\documentclass{aa}  

\usepackage{graphicx}
\graphicspath{{.}{figures/}}
\usepackage{txfonts}
\usepackage{mathptmx}      
\usepackage{xspace}   
\usepackage{gensymb}  
\usepackage{amssymb}  
\usepackage{amsmath}  

\usepackage{pgfpages} 
\usepackage{pgfplots}
\usepgfplotslibrary{external} 
\usetikzlibrary{pgfplots.external}
\usetikzlibrary{calc}
\usetikzlibrary{patterns}

\pgfplotsset{
  compat=1.4
  ,tick label style={font=\normalsize}
  ,ylabel absolute 
  ,label style={font=\normalsize}
  ,minor x tick num=3
  ,minor y tick num=4
  ,scaled ticks=false
  ,/pgf/number format/set thousands separator={\,}
}

\pgfkeys{/pgf/number format/.cd,fixed,precision=4}

\usepackage{lscape}

\usepackage{natbib}
\bibpunct{(}{)}{;}{a}{}{,} 

\begin{document}

\title{Multitechnique testing of the viscous decretion disk model}

   \subtitle{I. The stable and tenuous disk of the late-type Be star $\beta$~CMi\thanks{Based partly on observations from Ond\v rejov 2-m telescope, Czech Republic; partly on observations collected at the European Southern Observatory, Chile (Prop. No. 093.D-0571); as well as archival data from programs 072.D-0315, 082.D-0189, 084.C-0848, 085.C-0911, and 092.D-0311; partly on observations from APEX collected via CONICYT program C-092.F-9708A-2013, and partly on observations from CARMA collected via program c1100-2013a.}}

   \author{R.~Klement
          \inst{1}
          \and
          A.~C.~Carciofi\inst{2}
        \and
        Th.~Rivinius\inst{3}
        \and
        D.~Panoglou\inst{2}
        \and
        R.~G.~Vieira\inst{2}
        \and
        J.~E.~Bjorkman\inst{4}
        \and
        S.~\v Stefl\thanks{Deceased}
        \and
        C.~Tycner\inst{5}
        \and
        D.~M.~Faes\inst{2}
        \and
        D.~Kor\v c\' akov\' a\inst{1}
        \and
        A.~Müller\inst{3}
        \and
        R.~T.~Zavala\inst{6}
        \and
        M.~Cur\' e\inst{7}
}

   \institute{Astronomical Institute of Charles University, Charles University in Prague,
              V Hole\v sovi\v ck\'ach 2, 180 00  Prague 8\\
              \email{robertklement@gmail.com}
        \and
             Instituto de Astronomia, Geof\'isica e Ci\^encias Atmosf\'ericas, Universidade de S\~ao Paulo, Rua do Mat\~ao 1226, Cidade Universit\'aria, 05508-900, S\~ao Paulo, SP, Brazil
        \and
            European Organisation for Astronomical Research in the Southern Hemisphere, Casilla 19001, Santiago 19, Chile
        \and
                        Ritter Observatory, Department of Physics \& Astronomy, University of Toledo, Toledo, OH 43606, UsA
        \and
        Department of Physics, Central Michigan University,
  Mount Pleasant, MI 48859, USA 
        \and
        US Naval Observatory, Flagstaff Station, 10391
  W.~Naval Observatory Rd., Flagstaff, AZ 86001         
        \and
        Instituto de F\' isica y Astronom\' ia, Facultad de Ciencias, Universidad de Valpara\' iso, Casilla 5030, Valpara\' iso, Chile
                             }
   \date{}

 
  \abstract
   {The viscous decretion disk (VDD) model is able to explain most of the currently observable properties of the circumstellar disks of Be stars. However, more stringent tests, focusing on reproducing multitechnique observations of individual targets via physical modeling, are needed  to study the predictions of the VDD model under specific circumstances. In the case of  nearby, bright Be star $\beta$~CMi, these circumstances are a very stable low-density disk and a late-type (B8Ve) central star. 
   }
   {The aim is to  test the VDD model thoroughly, exploiting the full diagnostic potential of individual types of observations, in particular, to  constrain the poorly known structure of the outer disk if possible, and to test truncation effects caused by a possible binary companion using radio observations.}
   {We use the Monte Carlo radiative transfer code {\ttfamily HDUST} to produce model observables, which we compare with a very large set of multitechnique and multiwavelength observations that  include ultraviolet and optical spectra, photometry covering the interval between optical and radio wavelengths, optical polarimetry, and optical and near-IR (spectro)interferometry. 
}
   {A parametric VDD model with radial density exponent of $n$ = 3.5, which is the canonical value for isothermal flaring disks, is found to explain observables typically formed in the inner disk, while observables originating in the more extended parts favor a shallower, $n$ = 3.0, density falloff. Theoretical consequences of this finding are discussed and the outcomes are compared with the predictions of a fully self-consistent VDD model. Modeling of radio observations allowed for the first determination of the physical extent of a Be disk (35$^{+10}_{-5}$ stellar radii), which might be caused by a binary companion. Finally, polarization data allowed for an indirect measurement of the rotation rate of the star, which was found to be $W \gtrsim 0.98$, i.e., very close to critical.}
   {}


   \maketitle
%

\section{Introduction}
\label{intro}



The gaseous circumstellar disks around classical Be stars are special universal laboratories. Although Be stars have been known for 150 years \citep{secchi} and the idea of a flattened envelope has existed since 1931 \citep{struve}, consensus regarding the physical process governing the disk structure, the viscous decretion disk (VDD) model, has emerged only recently. 

The VDD model, first introduced by \citet{lee} and further developed by, e.g., \citet{bjorkman1997}, \citet{porter1999}, \citet{okazaki}, and \citet{bjorkman_carciofi_2005}, is now widely accepted as the best physical model for describing the circumstellar disks of Be stars. Among the growing evidence supporting the VDD model is the confirmation that the disks rotate in a Keplerian way \citep{meilland2007,kraus,wheel}, allowing for the identification of viscosity as the mechanism that makes the disk grow: Viscous torques transfer angular momentum from the base of the disk outward, thus allowing the gas particles to reach progressively wider orbits. Decretion disks share most of the physical characteristics with accretion disks known, e.g., from protostars, with their name referring to the outward direction of the mass flow. The mass gets injected into the decretion disk by an as of yet not unambiguously defined mechanism, most probably a combination of nonradial pulsations, fast rotation, and possibly small-scale magnetic fields \citep{review}.  


Stringent tests of the VDD model focusing on reproducing the observables of individual targets are however still very few in number. The VDD model implemented in the Monte Carlo radiative transfer code {\ttfamily HDUST} \citep{hdust1,hdust2} was used to reproduce the violet-to-red peak ratio ($V/R$) spectroscopic variability along with photometry, polarimetry, and near-infrared (IR) interferometry of $\zeta$ Tau, providing firm evidence that the $V/R$ oscillations are an effect of one-armed density waves in the disk \citep{ztau2}. Other successful {\ttfamily HDUST} applications include the visual light-curve modeling of the 2008 outburst of 28~CMa along with the first determination of the viscosity parameter \citep{viscosity}, and the modeling of visual and IR spectral energy distribution (SED) and visual spectropolarimetry of $\delta$~Sco \citep{dsco}. Further studies of individual stars include those of \citet{jones} and \citet{tycner2008} in which the {\ttfamily BEDISK} code \citep{sigutjones} was used. Circumstantial evidence supporting the VDD model is also provided, e.g., by \citet{haubois1}, who were able to reproduce observed light curves of Be stars with dynamical models; by \citet{silaj}, who reproduced generally well the observed H$\alpha$ profiles of 56 Be stars; and by \citet{touhami}, who were able to reproduce the statistical properties of the color excesses of a sample of 130 stars. The basic characteristics of the VDD model and  {\ttfamily HDUST} code are described in Sect. \ref{s3}. For an extensive review on classical Be stars and their disks, see \citet{review}.

When studying a complex astrophysical system, such as a star irradiating a surrounding circumstellar disk, it is crucial to combine different observational techniques, as each probes different aspects of the disk physics. The combination of  radiative transfer and disk dynamics  governs most observables, and their disentangling is the central goal of the modeling process. The most characteristic observables of Be stars are the often double-peaked hydrogen emission lines and the disk excess IR emission due to free-free and bound-free emission in the ionized circumstellar disk. Observed optical and near-IR radiation is partially linearly polarized as a result of scattering off free electrons in the inner disk. Medium- and high-resolution differential spectrointerferometry across an emission line, such as Br$\gamma,$ also shows patterns in visibilities and phases that are characteristic of circumstellar disks rotating in a Keplerian way. The combination of spectroscopic, photometric, polarimetric, and interferometric observations is therefore particularly strong for the case of circumstellar disks, removing some of the degeneracies arising from fitting single observations. 

The main goal of this study is to apply the VDD model to the case of the well-known Be star $\beta$~CMi and to test how well it handles arguably the largest set of multiwavelength and multitechnique observations of a Be star so far, thereby broadening the sample of detailed VDD tests applied on individual targets. The data set includes a compiled SED covering the ultraviolet (UV) to radio wavelengths, high-resolution spectroscopy of the emission lines, polarimetry, and high-resolution spectrointerferometry (see Sect.~\ref{s2}). 


Besides being nearby, bright and therefore often observed, $\beta$~CMi was chosen as the target for this study for two additional reasons. Firstly, the disk of $\beta$~CMi has been been present and stable for decades. \citet{slettebak} reported "no appreciable change" in line emission in the years 1950 to 1982 and the same holds for measurements from 1989 and 1991, which were published in \citet{hanuschik}. This implies that the disk does not go through major mass ejection and dissipation episodes often observed in many Be stars. The absence of high-amplitude $V/R$ variations is also evidence of no large-scale density fluctuations throughout the disk. The uncertainties arising from combining observations from different epochs should therefore be minimized. Secondly, $\beta$~CMi represents one of the first detailed VDD applications to a late-type Be star \citep[B8Ve,][]{abt}, thus enabling  testing of the universality of the VDD model throughout emission line stars of the spectral class B. Moreover, it is not entirely clear whether late-type Be stars differ qualitatively from the early-type stars with respect to, e.g., the observed variability (which seems to be lower for later types) and the rotation rate \citep[higher for later types, see Sects. 3.1 and 3.2 of][]{review}.

Probing the outer reaches of the disk by analyzing radio observations is another main goal of our research. For cases nonedge-on, the SED fluxes can be interpreted as the superposition of the photospheric and disk contributions. The free-free mechanism dominates the disk opacity in the near-IR and longward, and increases with wavelength approximately as $\propto\lambda^2$. Consequently, in a typical Be disk the central parts are optically thick along the line of sight and act like a pseudophotosphere, which longward of a density-dependent $\lambda_0$ grows in size according to the approximate relation $\overline{R}\propto\left({\lambda}/{\lambda_0}\right)^{0.41}$, where $\overline{R}$ is the pseudophotosphere radius \citep[][submitted]{vieira}. Therefore, different wavelengths probe distinct regions of the disk \citep[see, for instance, Fig. 2 of][]{review}. In this context, radio observations are of particular interest in studying the outer parts of the disk, and may eventually impose important constraints to the disk physical size. Radio observations were used to study the structure of the outer Be disks in the pioneering works of \citet{taylor}, \citet{waters}, and \citet{dougherty_radiovar}, however, since then no more radio studies of Be stars were published until very recently (see Sect.~\ref{radio_obs}). 

The binary nature of $\beta$~CMi is still an open question. It was suggested as a spectroscopic binary by \citet{jarad}; however, their data cover less than two cycles (considering their period of 218.5 days) and the spectral lines in the interval 3700 -- 4700~{\AA, which}  they used to determine the radial velocities, could be contaminated by disk emission. Their results for the radial velocities are therefore questionable and have not been confirmed. Recently, \citet{folsom} reported weak $V/R$ oscillations ($V/R$ = 0.98 -- 1.04) of the H$\alpha$ emission line with the period of 182.83 days discovered by a spectroscopic monitoring campaign in the years 2000 -- 2014. These variations may be interpreted as arising from disk asymmetries triggered by the tidal interaction between the disk and an unseen binary companion. In this case, the $V/R$ oscillations would be phase-locked with the binary orbit and therefore the $V/R$ period would be equal to the binary orbital period. 

Binary interaction is the only known mechanism that can truncate the disk interior to the photoevaporation radius. The so-called truncation radius represents a position in the disk where the tidal torque caused by the companion is in balance with the viscous torque. At this radius the disk is not simply cut off, but rather separated into two density regimes: Within the truncation radius the radial density decrease dependence can become shallower than an equivalent, isolated disk, and outside the truncation radius it becomes steeper, while the magnitude of this effect depends on the viscosity parameter \citep{okazaki2002,panoglou} and on the binary mass ratio (Sect.~\ref{binarity}). If the reported $V/R$ variations are indeed the effect of a binary orbit, radio observations might in principle be used to detect and quantify the truncation of the disk by the companion. In view of the possible change of the density structure with respect to steady viscous decretion caused by a binary truncation or a nonsteady decretion \citep{haubois1}, we investigate whether steady viscous decretion provides an adequate description or whether a more complicated density structure is necessary.

The last goal is to look for observable signatures of the disk nonisothermality, which is expected from theory. Be disks become nonisothermal because the inner, dense parts of the disk are optically thick to the photoionizing stellar radiation, causing an initial temperature decrease in the radial direction. As the density gets lower further from the star, the temperature rises to about 60\% of the stellar effective temperature and the disk becomes roughly isothermal. The temperature structure is expected to have an effect on the density structure of the disk, which then affects the observables \citep{hdust2}. 

In Sect. \ref{s2} the observations are described, while Sect. \ref{s3} briefly describes the adopted modeling procedure. In Sect. \ref{s4} the model predictions are presented and with their help the observations are interpreted. The conclusions follow.


\section{Observations}
\label{s2}

\subsection{Optical and IR photometry} 
\label{s2.1}

Standard Johnson \textit{UBVRI} photometry was taken from the catalog of \citet{ducati}. The \textit{UBVRI} photometry was dereddened with the reddening curve of \citet{cardelli} using the extinction coefficient $R_V = 3.1$ and the interstellar color excess $E[B-V] = 0.01$ adopted from the analysis of \citet{dougherty1994}. The overall reddening effects are found to be very weak (and negligible longward of 7000~\AA), as is also indicated by the very low extinction toward $\beta$~CMi \citep[$A_V$ = 0.05 according to][]{wheel}. 

IR photometry in filters \textit{JHKLM} was adopted from \citet{dougherty}. We also make use of the color-corrected IRAS measurements of $\beta$~CMi published in \citet{cotewaters}. The remaining IR photometry was taken from the published measurements from the AKARI/IRC mid-IR all-sky Survey \citep{ishihara}, from the Spitzer Space Telescope \citep[SST;][]{spitzer} and from the Wide-field Infrared Survey Explorer \citep[WISE;][]{wise}. Finally, we use the $N$-band flux calibrated spectrum ($R$ = 30) measured by the VLTI/MIDI interferometer. The compiled IR data set is listed in Table~\ref{t1}.

\begin{table}
\caption{\label{t1} The observed IR and radio fluxes.}
\centering
\begin{tabular}{ccc}
\hline\hline
$\lambda$ ($\mu$m)&Flux (Jy)& Instrument\\
\hline
 9   & 4.892 $\pm$ 0.005 & AKARI\\
 8 -- 13   & 3.31 $\pm$ 0.27\tablefootmark{a} & MIDI\\
 11.6   & 2.78 $\pm$ 0.03 & WISE\\
 12     & 3.21 $\pm$ 0.16 & IRAS\\     
 18  & 1.988 $\pm$ 0.061 & AKARI\\        
 22.1   & 1.58 $\pm$ 0.03 & WISE\\
 23.7   & 1.47 $\pm$ 0.01 & SST\\
 25     & 1.52 $\pm$ 0.15 & IRAS\\
 60     & 0.59 $\pm$ 0.06 & IRAS\\
 71.4   & 4.23e-1 $\pm$ 0.51e-1 & SST\\
 870    & 3.35e-2 $\pm$ 0.65e-2 & APEX/LABOCA\\
 1100   & 3.7e-2 $\pm$ 0.6e-2 & JCMT\\
 3265   & 9.6e-3 $\pm$ 0.6e-3 & CARMA\\
 20000  & 0.69e-3 $\pm$ 0.10e-3 & VLA\\
\hline
\end{tabular}
\tablefoot{\\
\tablefoottext{a}{The flux value is for the central $N$-band wavelength of 10.5~$\mu$m}
}
\end{table}

\subsection{Radio measurements}
\label{radio_obs}

Radio observations of Be stars have been very scarce throughout  decades past. However, recent years saw a great technological improvement in the instruments allowing for precise measurements even of very low fluxes. We briefly discuss the historic measurements and then describe our own from recent years.

Already in the late eighties, several Be stars including $\beta$~CMi were detected at 2~cm wavelength by the Very Large Array \citep[VLA;][]{taylor} and at mm wavelengths by the James Clerk Maxwell Telescope \citep[JCMT;][]{waters}. We use the published JCMT and VLA data in our analysis (see the respective references for the description of the data). To our knowledge, these had been the only radio observations of Be stars up until a few years ago, when 28~CMa was observed by the Atacama Pathfinder Experiment (APEX) millimeter telescope \citep{mm1} and $\delta$ Sco was observed by APEX and by the Combined Array for Research in Millimeter-wave Astronomy \citep[CARMA,][]{mm2}. More recently, we used the APEX bolometer camera LABOCA \citep{laboca} and the CARMA array \citep{carma} in its photometric mode to observe $\beta$~CMi. The sub-mm and radio observations are included in Table~\ref{t1}.

The APEX/LABOCA camera operates at the central frequency of 345~GHz, corresponding to the wavelength of 870~$\mu$m with a bandwidth of 150~$\mu$m. It consists of a 295-channel bolometer array laid out in concentric hexagons around a central channel. The sub-mm radiation is absorbed by a heat-sensitive semi-conductor, which measures temperature variations of a thin titanium film. Our measurement consists of five scans with a total integration time of 53.8~min. The data were reduced with the help of the free reduction software CRUSH\footnote{\url{http://www.submm.caltech.edu/~sharc/crush/index.html}} version 2.20-3 \citep{crush}.

The CARMA observations were executed with the 15-antenna subarray (nine 6.1-m and six 10.4-m diameter antennas) in the D configuration, in which the distances between the individual antennas are 11--150~m. The central frequency of the observations was $\sim$100 GHz ($\sim$3~mm) with the bandwidth of $\sim$500~MHz. The total integration time including the calibrators was 2.60 hours. The data were reduced using the MIRIAD software package\footnote{\url{http://carma.astro.umd.edu/miriad/}}.

\subsection{Spectroscopy}
\label{spec}

\begin{table*}
\caption{\label{t2}Spectroscopic data set.}
\centering
\begin{tabular}{cccccc}
\hline\hline
Date&Telescope&Instrument&No. of spectra&$R$&Spectral range (\AA)\\
\hline
1986-04-05 -- 1987-12-05 & IUE & IUE\_SWP\_HL & 4 & 18 000& 1150 -- 1975\\
1987-02-10 & IUE & IUE\_LWP\_HL & 1 & 13 000& 1850 -- 3350\\
1995-02-08 -- 2000-02-12 & PBO 36'' telescope & HPOL & 3 & 600 & 3200 -10500\\
2001-11-16 -- 2007-01-08 & Ond\v rejov 2~m & coud\'e spectr. & 11 & 12 500& 6300 -- 6700\\
2004-08-01 -- 2004-09-01 & ESO - La Silla 2.2~m & FEROS & 2 & 48 000& 3600 -- 9200\\
2007-01-08 -- 2014-02-23 & Ond\v rejov 2~m & coud\'e spectr. & 9 & 18 500& 4760 -- 5000 \\ 
2008-01-26 & Meade LX200 & LHIRES & 1 & 17 000& 6300 -- 6700\\
2009-02-14 & C11 & LHIRES & 1 & 17 000& 6300 -- 6700\\
2009-12-31 -- 2010-04-23 & VLTI & AMBER & 2 & 12 000& 21580 -- 21760\\
2010-01-16 & Takahashi CN212 & LHIRES & 1 & 15 000& 6300 -- 6700\\
2010-10-29 & C11 & LHIRES & 1 & 17 000& 6300 -- 6700\\
2011-10-20 -- 2013-09-26 & OPD 1.6~m & ECass & 9 & 5 000 & 6120 -- 6880\\2011-11-05 -- 2011-11-11 & CFHT 3.6~m & ESPaDOnS & 80 & 68 000 & 3700 - 9000\\
2013-12-09 -- 2014-11-18 & OPD 1.6~m & MUSICOS & 3 & 30 000 & 5020 -- 9020\\
2013-02-08 -- 2014-02-23 & Ond\v rejov 2~m & coud\'e spectr. & 7 & 12 500& 6300 -- 6700\\
2013-11-09 -- 2014-02-23 & Ond\v rejov 2~m & coud\'e spectr. & 3 & 16 500& 4280 -- 4500\\ 
\hline
\end{tabular}
\end{table*}

The spectral coverage of the compiled SED is extended into ultraviolet (UV) wavelengths by including spectra measured by the International Ultraviolet Explorer (IUE)\footnote{\url{https://archive.stsci.edu/iue/}}. We observed $\beta$~CMi  by IUE five times in high dispersion mode with the total spectral coverage of 1150 -- 3350~{\AA}. The spectra were averaged and then dereddened with the same approach as the optical photometry with the reddening curve updated for the near-UV according to \citet{odonnell}. Strong reddening is usually indicated by an interstellar absorption feature around 2200~{\AA}. The fact that this feature is absent from the observed spectra is a further indication of low extinction toward $\beta$~CMi and overall weak reddening effects.

In the optical region, the Johnson \textit{UBVRI} photometry is complemented with several flux-calibrated low-resolution spectra covering the interval 3200 -- 10500~{\AA} measured by the HPOL spectropolarimeter\footnote{\url{http://archive.stsci.edu/hpol/}} mounted on the Pine Bluff Observatory (PBO) 36'' telescope. As the absolute flux calibration of the spectra is not reliable, however, only the shape of the spectra was used. The individual measurements were scaled to the $V$-band magnitude before they were averaged and dereddened. 

\begin{figure}
   \centering
   \includegraphics[width=\hsize]{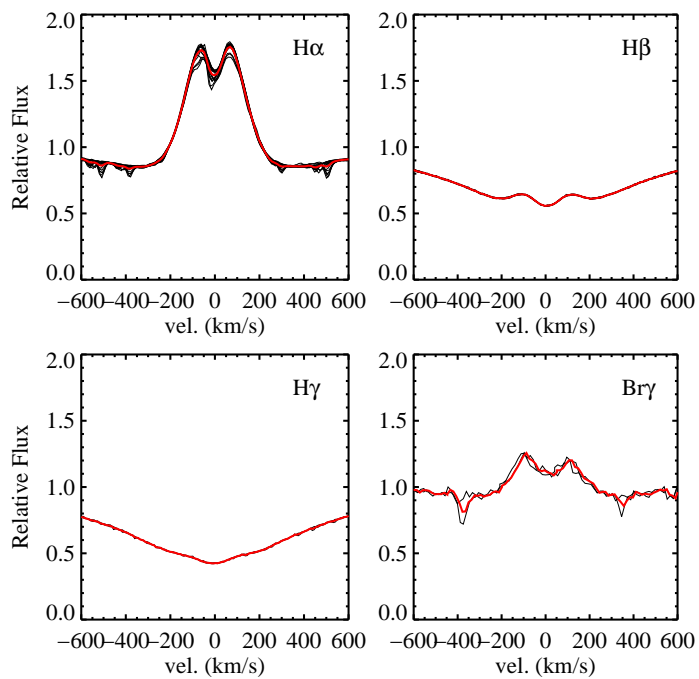}
      \caption{Balmer line profiles from Ond\v rejov and Br$\gamma$ profile from AMBER (black) overplotted with their averages (red), which were subsequently used in the modeling process. Telluric features can be seen in most of the H$\alpha$ and Br$\gamma$ profiles.
              }
         \label{plotlines}
   \end{figure}

The spectroscopic observations of hydrogen Balmer emission lines ($R$ = 5000 -- 30000) in the 2001 -- 2014 interval were carried out using the slit spectrograph mounted at the coud\'e focus of Ond\v rejov 2~m telescope in the Czech Republic and the Cassegrain (ECass) and MUSICOS spectrographs at the 1.6~m telescope at the Observat\' orio Pico dos Dias (OPD) in Brazil. The gaps in the time coverage were filled in by several amateur spectra from the BeSS database\footnote{\url{http://basebe.obspm.fr/basebe/}}. The Balmer lines are complemented with Br$\gamma$ profiles extracted from two sets of AMBER interferometric measurements (Sect.~\ref{interfer}). Overall, the line profiles show only small-scale temporal variability, although some show a weak $V/R$ assymetry, in line with the results of \citet{folsom}. In Fig.~\ref{plotlines} we show the Ond\v rejov and AMBER line profiles, the averages of which were used for the modeling process.

Two spectra from the high-resolution ($R$ = 48 000) fiber-fed spectrograph FEROS \citep{feros}, mounted at the 2.2~m ESO telescope at the La Silla observatory, Chile and a set of spectroscopic measurements from the ESPaDOnS\footnote{\url{http://www.ast.obs-mip.fr/projets/espadons/espadons.html}} echelle spectrograph ($R$ = 68 000), installed at the 3.6~m Canada-France-Hawaii Telescope (CFHT) in Hawaii, USA, allowed us to investigate the whole spectrum in the 3600 -- 9200~{\AA} range. 

The information on the spectroscopic data set is summarized in Table~\ref{t2}.

\subsection{Linear polarimetry}

$\beta$~CMi was observed by the HPOL Spectropolarimeter mounted on the PBO 36'' telescope six times during the years 1991 -- 2005. The first measurement was done with a Reticon dual array detector, while a CCD detector was used for the remaining scans. The wavelength coverage of the CCD observations is 3200 – 10500~{\AA} with a spectral bin size $\sim$10~\AA. 

Recent broadband polarimetric measurements in the Johnson filters \textit{BVRI} were obtained on five observing nights during the years 2011 -- 2014 at 1.6~m telescope of the OPD observatory in Brazil (Table~\ref{onlopd_polarimetry}, available electronically only). For the instrument description, observing setup and calibration and reduction procedures see Sect.~2 of \citet{carciofi2007} and references therein.

\onltab{
\begin{table*}
\caption{OPD/LNA broadband polarimetry.}\label{onlopd_polarimetry}
\centering
\begin{tabular}{cccc}
\hline  \hline
Date & Johnson filter & Polarization level (\%) & Position Angle (\degree) \\
\hline
2011-10-20 & B & 0.039 $\pm$ 0.003 & 73.7 $\pm$ 2.0 \\
2011-10-20 & V & 0.022 $\pm$ 0.015 & 51.5 $\pm$ 20.1 \\
2011-10-20 & R & 0.050 $\pm$ 0.004 & 46.1 $\pm$ 2.3 \\
2011-10-20 & I & 0.071 $\pm$ 0.019 & 57.0 $\pm$ 7.8 \\
2012-04-04 & V & 0.027 $\pm$ 0.018 & 38.8 $\pm$ 18.6 \\
2012-04-04 & R & 0.027 $\pm$ 0.015 & 37.8 $\pm$ 15.9 \\
2012-04-04 & I & 0.049 $\pm$ 0.015 & 43.3 $\pm$ 8.8 \\
2013-04-24 & B & 0.048 $\pm$ 0.076 & 37.6 $\pm$ 46.0 \\
2013-04-24 & V & 0.038 $\pm$ 0.039 & 54.4 $\pm$ 29.1 \\
2013-11-13 & B & 0.059 $\pm$ 0.064 & 43.3 $\pm$ 31.3 \\
2013-11-13 & V & 0.060 $\pm$ 0.035 & 55.2 $\pm$ 16.7 \\
2013-11-13 & R & 0.053 $\pm$ 0.024 & 41.8 $\pm$ 13.2 \\
2013-11-13 & I & 0.078 $\pm$ 0.013 & 44.8 $\pm$ 4.8 \\
2014-03-09 & B & 0.096 $\pm$ 0.143 & 67.0 $\pm$ 42.9 \\
2014-03-09 & V & 0.042 $\pm$ 0.023 & 60.4 $\pm$ 15.9 \\
2014-03-09 & R & 0.066 $\pm$ 0.024 & 51.8 $\pm$ 10.4 \\
2014-03-09 & I & 0.038 $\pm$ 0.021 & 45.2 $\pm$ 15.6 \\
\hline
\end{tabular}
\end{table*}
}

As discussed in Sects.~\ref{s2.1} and \ref{spec}, the interstellar extinction toward $\beta$~CMi is almost negligible, therefore, we perform no corrections for the interstellar polarization.

\subsection{Interferometry}
\label{interfer}

\begin{table*}
\caption{\label{t3}Interferometric observations from VLTI and CHARA.}
\centering
\begin{tabular}{cccccccc}
\hline\hline
Date&Instrument&Spectral band& $R$ & {NP} \tablefootmark{a}&Telescope configuration \\
\hline
2008-12-02 & MIRC & $H$ & 35 & 1 & S1-E1-W1-W2 \\
2008-12-17 & MIDI & $N$ & 30 & 1 & UT1-UT4 \\
2009-01-11 & MIDI & $N$ & 30 & 1 & UT1-UT4 \\
2009-11-10 & MIRC & $H$ & 35 & 3 & S1-E1-W1-W2 \\
2009-12-31 & AMBER & $K$ & 12 000 & 1 & UT1-UT2-UT3 \\
2010-04-23 & AMBER & $K$ & 12 000 & 1 & K0-G1-A0 \\
2010-10-30 & CLIMB & $K$' & - & 2 & S2-W1-W2 \\
2010-11-02 & MIRC & $H$ & 35 & 3 & S1-E1-W1-W2 \\
2010-11-03 & MIRC & $H$ & 35 & 2 & S2-E2-W1-W1 \\
2010-12-02 & CLIMB & $K$' & - & 1 & S1-E2-W1 \\
2010-12-03 & CLIMB & $K$' & - & 4 & S2-E2-W2 \\
2010-12-14 & MIRC & $H$ & 35 & 1 & S1-E2-W1-W2 \\
2013-10-21 & MIDI & $N$ & 30 & 1 & UT1-UT4 \\
2014-03-06 & FSU-A & $K$ & - & 2 & J3-K0, G1-A1 \\
2014-03-07 & FSU-A & $K$ & - & 1 & K0-A1 \\
\hline
\end{tabular}
\tablefoot{\\
\tablefoottext{a}{Number of pointings obtained on the science star.}}
\end{table*}

\onllongtab{
\begin{longtable}{ccccc}
\caption{\label{t4} Calibrated NPOI H$\alpha$ squared visibilities of $\beta$~CMi.}\\
\hline\hline
Julian Date & Spatial Frequency $u$ & Spatial Frequency $v$ &   &  \\
(JD - 2 450 000) & (10$^6$ cycles rad$^{-1}$) & (10$^6$ cycles rad$^{-1}$) & $V^2$ & Baseline\\
\hline
\endfirsthead
\caption{Continued.} \\
\hline
Julian Date & Spatial Frequency $u$ & Spatial Frequency $v$ &   &  \\
(JD - 2 450 000) & (10$^6$ cycles rad$^{-1}$) & (10$^6$ cycles rad$^{-1}$) & $V^2$ & Baseline\\
\hline
\endhead
\hline
\endfoot
\hline
\endlastfoot
5519.998  &  $  27.398$  &  $  -8.902$  &  0.946 $\pm$ 0.022 & AC-AE \\
5519.998  &  $  58.731$  &  $  39.487$  &  0.770 $\pm$ 0.042 & AC-W7 \\
5519.998  &  $  86.129$  &  $  30.584$  &  0.677 $\pm$ 0.039 & AE-W7 \\
5519.998  &  $  50.868$  &  $  -9.168$  &  0.862 $\pm$ 0.048 & AC-E6 \\
5519.998  &  $ 109.572$  &  $  30.301$  &  0.537 $\pm$ 0.027 & E6-W7 \\
5519.998  &  $  58.704$  &  $  39.469$  &  0.662 $\pm$ 0.032 & AC-W7 \\
5531.940  &  $  27.106$  &  $  -9.530$  &  0.863 $\pm$ 0.041 & AC-AE \\
5531.940  &  $  64.257$  &  $  38.071$  &  0.786 $\pm$ 0.029 & AC-W7 \\
\end{longtable}
}

The interferometric observations were obtained at three interferometric facilities: Very Large Telescope Interferometer (VLTI) at the Paranal observatory in Chile, Center for High Angular Resolution Astronomy (CHARA) at the Mount Wilson Observatory in California, USA, and the Navy Precision Optical Interferometer (NPOI) located in Arizona, USA. The VLTI instruments include the 2-telescope beam combiners PRIMA FSU-A \citep{prima} and MIDI \citep{midi} and the 3-telescope beam combiner instrument AMBER \citep{amber}. We also observed $\beta$~CMi for two nights (May 1 -- May 3, 2014) with the 4-telescope beam combiner PIONIER \citep{pionier}, but these measurements were mostly lost due to bad weather conditions. The CHARA instruments used are the 4 to 6-telescope beam combiner MIRC \citep{mirc} and a 3-beam combiner CLIMB \citep{climb}. The AMBER, MIRC, and CLIMB data were already analyzed by \citet{kraus} and part of the MIDI observations was published in \citet{meilland_midi}. The observation logs for interferometric observations from VLTI and CHARA are listed in Table~\ref{t3} and the full NPOI data set is given in Table~\ref{t4} (available electronically only).

The former PRIMA (Phase-Referenced Imaging and Micro-arcsecond Astrometry) facility at VLTI consisted of two identical fringe sensor units (called FSU-A and FSU-B) but only FSU-A was publicly offered for observations. The FSU-A  operates in the $K$-band between 2.0\,$\mu$m and 2.5\,$\mu$m. The light is chromatically dispersed over five spectral pixels. Three visibility measurements were obtained with pairs of 1.8~m Auxiliary Telescopes (ATs). For more details on the observing setup and the reduction procedure, see Sect.~4 of \citet{muller}.

The MIDI data consist of three measurements executed with pairs of 8.2~m Unit Telescopes (UTs) with spectral dispersion $R$ = 30 across the $N$-band (8 -- 13~$\mu$m). For more details on the data acquisition and reduction, see \citet{meilland_midi} and references therein.

The AMBER observations were obtained in the high spectral resolution mode (HR mode, $R$ = 12 000) in the $K$-band spectral region centered around the hydrogen Br$\gamma$ line ($\lambda_{\text{Br}\gamma}^{\text{vacuum}} = 2.166078~\mu$m). The pointing from December 31, 2009 was obtained with three 8.2~m UTs, while for the observation from April 23, 2010 three 1.8~m ATs were used. For our purposes, we rereduced the data using the amdlib V3.0 data reduction software \citep{tatulli, chelli}. As absolute calibration of the AMBER data is not reliable, we normalized the continuum region and used the differential visibilities and phases for our analysis. As the measurement error for all data points we used the standard deviation computed from the continuum region. For more details on the observing setup and the reduction procedure see Sect.~2 of \citet{kraus}. 

The observations from MIRC were obtained with low spectral dispersion, $R$ = 35, in the $H$-band continuum. The light from four 1~m telescopes was combined with the baseline lengths reaching 330~m. The CLIMB observations include eight pointings in the $K$'-band continuum. The observing setup and the reduction process are detailed in Sect.~2 of \citet{kraus}. For the $uv$-plane coverage of the CHARA measurements as compared to AMBER coverage, see their Fig.~1.

The NPOI is a long-baseline, multielement, optical interferometer that can combine light beams from up to six elements simultaneously \citep{npoi}. Before the fringe contrast is recorded between the beams, the light combined at the beam combiner is dispersed using a prism onto a lenslet array connected by optical fibers to a series of photon-counting avalanche photodiodes, which results in data stream from 16 spectral channels covering the wavelength range of 560 -- 870~nm.  For B-type stars with an H$\alpha$ emission line, it allows interference fringe contrast (in the form of squared visibility) to be recorded for a single 15-nm wide spectral channel that contains the entire emission line from the circumstellar region (we refer to this as H$\alpha$ channel when the 656.28~nm line is centered in the spectral channel).

The NPOI observations of $\beta$~CMi utilized in this study were obtained on 40 nights from November 2010 to April 2011. The entire data set resulted in 720 unique squared visibility measurements from the H$\alpha$ channel. The processing of NPOI data was conducted using a custom suite of programs developed specifically for the instrument known as the Optical Interferometer Script Data Reduction package. The processing of the raw fringe data frames follows the procedures outlined in \citet{hummel1998} with additional bias corrections using off-fringe measurements described in \citet{hummel2003}. Furthermore, in the case of H$\alpha$-emitting sources, where the signal of interest is confined to a single spectral channel per baseline, the additional step of removal of small channel-to-channel variations has also been conducted as described in \citet{tycner2006} utilizing a calibrator star $\lambda$~Gem (FK5 277, HR 2763). The last step of calibration of the H$\alpha$ squared visibilities involved calibration with respect to the continuum channels, which  are assumed to be represented by a uniform disk model with a known angular diameter \citep{tycner2003}. In the case of $\beta$~CMi, we adopted an angular diameter of the central star of 0.675 mas, which was obtained based on a Hipparcos distance of 49.6~pc \citep{hip}, the best-fit value for stellar polar radius ($R_\text{p}$) of 2.8~R$_\odot$ based on UV spectrum fitting (see Sect.~\ref{uv}) and a multiplicative factor of 1.29, which accounts for the average effect due to the rapid stellar rotation. \footnote{To account for the rapid stellar spin, we take the average between equatorial radius $R_\text{e}$ = 1.49~$R_\text{p}$ (see Table~\ref{derived}) and the projection of the equatorial radius due to inclination $i$ of 43{\degree} (see Sect.~\ref{amber}), i.e., $\cos(43) = 0.731$ resulting in average effective radius of 1.29~$R_\text{p}$.}



\section{Model description}
\label{s3}

We modeled $\beta$~CMi  as a combination of a fast-rotating central star surrounded by a viscous decretion disk with the help of the computer code {\ttfamily HDUST}. The code {\ttfamily HDUST} uses the Monte Carlo method to solve the radiative transfer, radiative equilibrium, and statistical equilibrium in 3D geometry for arbitrary density and velocity distributions assuming pure hydrogen composition. For details on the {\ttfamily HDUST} code and the hydrodynamics of a VDD; see \citet{hdust1,hdust2} and \citet{carciofi2011}.

\subsection{Central star}

The code {\ttfamily  HDUST} takes into account both the geometrical deformation of the star due to fast rotation and the latitude-dependent flux from the stellar surface due to gravity darkening. The stellar surface itself is divided into a number of latitude bins each with its own local gravity and temperature with a corresponding Kurucz \citep{kurucz} synthetic spectrum assigned to it.

A rotating star is fully described by five parameters: mass $M$, polar radius $R_\text{p}$, luminosity $L$, rotation rate $W$, and gravity darkening exponent $\beta$. The star is a geometrically deformed isopotential surface in the Roche approximation based on the rotation rate $W$, defined as $W = v_{\text{rot}}/v_{\text{orb}}$, where $v_{\text{rot}}$ is the rotational velocity at the stellar equator and $v_{\text{orb}}$ is the Keplerian velocity of orbiting material just above the stellar equator \citep[Sect. 2.3.1 of][]{review}. In the case of critical rotation, the parameter $W$ becomes unity. The gravity darkening law is adopted in the classical form $T_{\text{eff}} \propto g^{\beta}$ \citep{vonzeipel}. 

Two additional geometrical parameters are needed  to be able to compare the model with observations: The distance $d$ and the inclination angle $i$ of the spin axis of the star with respect to the observer.

\subsection{The disk}
\label{s3.2}

The VDD model in which turbulent viscosity is responsible for the disk growth and transport of the angular momentum outward has so far been able to explain most observational aspects of the Be star disks \citep{bjorkman}. 

The VDD is flaring with the following dependence of the scale height $H$ on the distance $r$ from the stellar surface, if one assumes hydrostatic equilibrium and an isothermal disk
\begin{equation}
\label{sch}
H(r) = H_0 \left(\frac{r}{R_\text{e}}\right)^{3/2},
\end{equation}
where $R_\text{e}$ is the equatorial radius of the rotationally deformed star and $H_0 = c_s v_\text{orb}^{-1} R_\text{e}$ is the scale height at the base of the disk. $c_\text{s} = [(k_\text{B} T_\text{k})/(\mu m_\text{H})]^{1/2}$ is the isothermal sound speed, $k_\text{B}$ is the Boltzmann constant, $T_\text{k}$ is the gas kinetic temperature, $\mu$ is the mean molecular weight, and $m_\text{H}$ is the mass of hydrogen atom. \citet{haubois1} studied the hydrodynamics of Be disks and explored how the disk surface density evolves with time in response to varying disk feeding rates. Even though a dynamically active disk can have a quite complex structure, a much simpler situation arises when the disk is fed at a constant rate for a sufficiently long time. In this case, the surface density structure of a VDD under the assumption of an isothermal gas and purely Keplerian circular orbits can be described by
\begin{equation}
\label{eq}
\Sigma (r) = \frac{\dot{M} v_\text{orb} R_\text{e}^{1/2}}{3\pi \alpha c_\text{s}^2 r^{3/2}} \left[(R_0/r)^{1/2}-1\right],
\end{equation}
where $\alpha$ is the viscosity parameter (assumed to be constant throughout the disk), $\dot{M}$ is the mass decretion rate, $r$ is the distance from the stellar equator, and $R_0$ is an integration constant connected with the physical size of the disk \citep{bjorkman_carciofi_2005}. For $r \ll R_0$, the dependence of $\Sigma(r)$ becomes a power-law $\Sigma(r) \propto r^{-2}$. The disk is in hydrostatic equilibrium in the vertical direction, which in the isothermal case corresponds to Gaussian distribution. The falloff of the volume density is then
\begin{equation}
\rho(r,z) = \frac{\Sigma(r)}{\sqrt{2\pi} H(r)} \text{exp} \left[-\frac{1}{2}\left(\frac{z}{H(r)}\right)^2\right],
\end{equation}
therefore
\begin{equation}
\rho \propto \Sigma / H \propto r^{-n}
,\end{equation}
with the radial density exponent $n$ equal to 3.5, however, the disk is not radially isothermal. Initially the temperature falls with radius to a minimum that occurs in the dense inner parts of the disk. The location of the minimum depends on the disk density; for higher densities it shifts outward. The temperature minimum results in a local change of the radial density falloff exponent $n$ and of the flaring coefficient \citep{hdust2}.

Numerical solutions \citep[e.g., ][]{okazaki2002} show that the disk is separated into two parts: a subsonic inner part (radial velocity $v_\text{r} \ll c_\text{s}$), where the outflow is driven by viscosity; and a transonic outer part ($v_\text{r} \gtrsim c_\text{s}$), where the outflow is driven by the gas pressure. The photoevaporation radius where the transition occurs is the so-called transonic critical radius $R_\text{c}$, which is on the order of hundreds of stellar radii \citep{krticka}. However, what is considered to be the outer radius of a Be disk  depends on whether the star is in a binary system or not. If the star is in a binary system, the disk may be truncated at the tidal radius, which can be a complicated function of the binary parameters and of the viscosity parameter $\alpha$ \citep{panoglou}. For isolated stars or well-detached binaries, however, we expect $R_\text{out} \approx R_\text{c}$. Radio observations have to be used to investigate these outer parts of the disk.

\begin{table*}
\caption{\label{t6}Characteristics and free parameters of the adopted VDD models.}
\centering
\begin{tabular}{ccc}
\hline\hline
Model&Characteristics&Free parameters\\ \hline
\textit{Parametric} & power-law radial density and vertical scale height structure& $\rho_0$, n, $H_0$ ($T_\text{k}$), $R_\text{out}$\\
\textit{Self-consistent} & full solution of the viscous decretion & $\dot{M}/\alpha$, $R_0$, $R_\text{out}$ \\
\hline
\end{tabular}
\end{table*}

\subsection{Modeling procedure}
\label{procedure}

For the present study, we have adopted two VDD models for the disk. First we start with a simplified analytical power-law model followed by the fully self-consistent solution of steady viscous decretion. As the latter model modifies the prescribed density structure according to the temperature solution, comparing the results of the two models allows us to test how the disk nonisothermality affects the observables. Improvement of the global fit to the observations after introducing the full VDD solution would then represent an indirect observational proof of the structural changes due to the disk nonisothermality.

The first  model, which we dub \textit{\textup{parametric model}}, is based on the isothermal VDD density structure for which the vertical distribution is Gaussian and the radial distribution is a power law in the form $\rho \propto \rho_0 r^{-n}$, where $\rho_0$ is the density at the base of the disk. Although this corresponds to the theoretical density structure of an isothermal disk (when $n$ = 3.5), the disk itself is not isothermal as its temperature is calculated in radiative equilibrium by {\ttfamily HDUST}. Values of $\rho_0$ were explored in the interval 1~$\times$~10$^{-12}$~--~1~$\times$~10$^{-10}$~g\,cm$^{-3}$, which is typical for Be disks. The scale height at the base of the disk $H_0$ is an additional parameter, which is set by the sound speed $c_\text{s}$ and the disk orbital speed $v_\text{orb}$ (see Eq.~\ref{sch}). As $v_\text{orb}$ is given by $M$, $R_\text{p}$, and $W$, only a proper choice of the kinetic temperature $T_\text{k}$ (to calculate $c_\text{s}$) is needed to determine the scale height $H_0$. We chose the mass-averaged temperature of the disk to represent $T_\text{k}$.

Varying the density falloff exponent $n$ allows us to explore a standard, isolated disk with isothermal density structure ($n$ = 3.5) as well as a disk truncated by a binary companion, in which the parts inward of the truncation radius may attain a shallower density profile ($n$ $<$ 3.5). Density profiles associated with a disk that has not yet reached a quasi-steady-state configuration \citep[$n$ $>$ 3.5, see][]{porter1999, haubois1} were also explored (although this case is unlikely for $\beta$~CMi, as its disk has been present and stable for a long period already). In our modeling, $n$ was allowed to vary in the interval 3.0~--~4.0.


The second, \textit{\textup{self-consistent model}} calculates the full solution of steady viscous decretion following \citet{hdust2}. In this case, the radial and vertical density structures are solved to become consistent with the temperature solution and therefore the parameters $n$ and $H_0$ are no longer needed. Instead of $\rho_0$ describing the base density, the ratio of the mass decretion rate and the viscosity parameter $\dot{M}/\alpha$ needs to be specified (see Eq. \ref{eq}). The $\dot{M}/\alpha$ parameter was chosen to correspond to the $\rho_0$ of the parametric model so that the resulting mass of the disk would remain approximately constant, thus allowing for a direct comparison between the models. The observables we study are only sensitive  to the density scale of the disk, and do not carry any information about its radial velocity because it is much smaller than the sound speed. For this reason, the mass-loss rate  $\dot{M}$ cannot be determined uniquely; we can only determine $\rho_0$ or equivalently the ratio $\dot{M}/\alpha$. Finally, there is an additional parameter in the form of the integration constant $R_0$, which was assumed to  correspond roughly to the critical radius $R_\text{c}$ given by the approximate relation of \citet{krticka}, i.e., 
\begin{equation}
\frac{R_\text{c}}{R_\text{e}} = \frac{3}{10} \left(\frac{v_\text{orb}}{c_s}\right)^2.
\end{equation}
The parameter $R_\text{c}$, calculated from the best-fit model values, is $\sim$450~$R_\text{e}$.


In both models, we also need to specify the disk outer radius $R_\text{out}$, which in the modeling context represents the outer edge of the grid used for the Monte Carlo simulation, beyond which matter is absent. This parameter allows us to explore possible truncation of the disk by a binary companion. 

The basic characteristics and free parameters of the models are listed in Table~\ref{t6}.

\begin{table*}[htb]
\caption{\label{t5}Model parameters}
\centering
\begin{tabular}{ccccc}
\hline\hline
Parameter&Best-fit value&Type&Adopted range&Ref.\tablefootmark{a}\\
\hline
&&Stellar parameters&\\
\hline
Spectral type & B8Ve & & &  2\\
$R_\text{p}$ & 2.8~$R_{\odot}$& free & 2.4 -- 3.6~$R_{\odot}$& 1, 3, 4 \\$M$ & 3.5~$M_{\odot}$ & fixed & & 4, 5\\
$L$ & 185~$L_{\odot}$ & free & 160 -- 240~$L_{\odot}$& 1, 3, 4, 5\\
$W$ & $\gtrsim 0.98$ & free & 0.7 -- 1.0 & 1, 5\\ 
$\beta$ & 0.1367$^{+0.0025}_{-0.0013}$ & function of $W$ & & 6\\
\hline
&&Disk parameters - \textit{parametric model}&\\
\hline
$\rho_0$ & 2.0 $\times$ 10$^{-12}$~g$\cdot$cm$^{-3}$& free & 1.0 $\times$ 10$^{-12}$ -- 1.0 $\times$ 10$^{-10}$~g$\cdot$cm$^{-3}$ & 1, 3\\
$n$ & 3.0 -- 3.5 & free & 3.0 -- 4.0 & 1\\
$T_\text{k}$\tablefootmark{b} &  6500 K & fixed &  & 1\\
$R_{\text{out}}$ & 35$^{+10}_{-5}$~$R_\text{e}$ & free & 10 -- 100~$R_\text{e}$ & 1\\
\hline
&&Disk parameters - \textit{self-consistent model}&\\
\hline
$\dot{M}/\alpha$& 1.88 $\times$ 10$^{-12}$~$M_{\odot}$\,yr$^{-1}$& free & 1.0 $\times$ 10$^{-12}$ -- 1.0 $\times$ 10$^{-11}$& 1\\
$R_0$ & 450~$R_\text{e}$& fixed & & 7\\
$R_{\text{out}}$ & $>$100~$R_\text{e}$ & free & 10 -- 200~$R_\text{e}$ & 1\\
\hline
&&Geometrical parameters&\\
\hline
$i$ & 43$^{+3\degree}_{-2\degree}$ & free & 35 -- 50\degree & 1, 3, 4\\
$d$ & 49.6 pc & fixed & & 8\\
$PA$\tablefootmark{c} & 133$^{+4\degree}_{-3\degree}$ & free & 0 -- 360\degree & 1\\
\end{tabular}
\tablefoot{\\
\tablefoottext{a}{References: 1~-~this work; 2~-~\citet{abt}; 3~-~\citet{wheel}; 4~-~\citet{kraus}; 5~-~\citet{saio}; 6~-~\citet{lara}; 7~-~\citet{krticka}; 8~-~\citet{hip}.}
\tablefoottext{b}{Assumed to be equal to the mass-averaged temperature of the disk.}
\tablefoottext{c}{The projected position angle of the semimajor axis of the disk measured from north to east.}
}
\end{table*}

\subsection{Initial estimates for the central star parameters}

$\beta$~CMi has already been a subject of several detailed studies focusing on different observational and physical aspects. $\beta$~CMi was previously studied by \citet{saio}, whose analysis of MOST satellite observations led to the first detection of nonradial pulsations in a late-type Be star with evidence for an almost critical rotation. Elsewhere, high-resolution spectrointerferometry and spectroscopy of $\beta$~CMi was used to provide evidence of Keplerian rotation of the disk \citep{kraus,wheel}. The central star parameters adopted in these studies were used as initial constraints for our modeling.

The value of $M$ = 3.5~$M_\odot$ was adopted in accordance with individual literature estimates \citep{kraus,saio} and stellar evolution models for the spectral type B8. Although $M$ influences the scale height of the disk as well as the critical rotation rate, changing it over the interval 3 -- 4~$M_\odot$ has little overall influence on the observables, and it was therefore kept fixed at the value of 3.5~$M_\odot$. 

We explored the possible values of $R_\text{p}$ and $L$ in the intervals 2.4 -- 3.6~$R_\odot$ and 160 -- 240~$L_\odot$, respectively. The rotation rate $W$ was assumed to be 0.7 -- 1.0, corresponding to the observed rotation rates of Be stars \citep[Sect. 3.1 of][]{review}. 

Recent interferometric observations as well as theoretical developments indicate that the gravity darkening law in the classical form $T_{\text{eff}} \propto g^{\beta}$ with $\beta = 0.25$ is valid only in the limit of slow rotators, while in rapidly rotating stars $\beta$ can be significantly lower. The recent model of \citet{lara} in which the barotropic assumption (pressure dependent only on density) is relaxed is in general agreement with six interferometric measurements of the $\beta$ parameter in rapid rotators, as is shown in Fig.~13 of \citet{domiciano}. In this work we treat the gravity darkening exponent $\beta$ as a function of $W,$ according to the model of \citet{lara}. However,  in their model, gravity darkening is in fact not described by a power law (see their Eq. 31).

\subsection{Geometrical parameters}

The distance $d$ to $\beta$~CMi is 49.6 $\pm$ 0.5~pc as measured by the Hipparcos satellite \citep{hip}. As varying $d$ in such a low uncertainty interval has little effect on the model observables, $d$ was kept fixed at the value of 49.6~pc. The inclination angle $i$ is clearly of an intermediate value with the most precise estimate from modeling of AMBER observations being 38.5 $\pm$ 1\degree \citep{kraus}. We explored the possible values of $i$ in the interval of 35\degree -- 50{\degree}. 


\section{Results}
\label{s4}

In this Sect. the ability of the VDD model, in the two particular formalisms adopted (\textit{\textup{parametric}} vs. \textit{\textup{self-consistent}}), to reproduce individual observations is detailed. 

The fitting procedure was executed in the following way: First, the almost purely photospheric UV spectrum was used to constrain the central star parameters $R_\text{p}$ and $L$. Next, the SED from IR to radio wavelengths, where the disk significantly contributes to the observed flux, was used to determine the disk parameters $\rho_0$ and $n$ (or $\dot{M}/\alpha$). Finally, the spectral slope of the linear polarimetry allowed us to determine the rotation rate $W$ (and $\beta$), while we used the interferometric shape extracted from the AMBER observations  to constrain the geometrical parameter $i$. The resulting best-fit parameters for both models are listed in Table~\ref{t5} and the derived stellar parameters are listed in Table~\ref{derived}.

\begin{figure}
   \centering
      \includegraphics[width=\hsize]{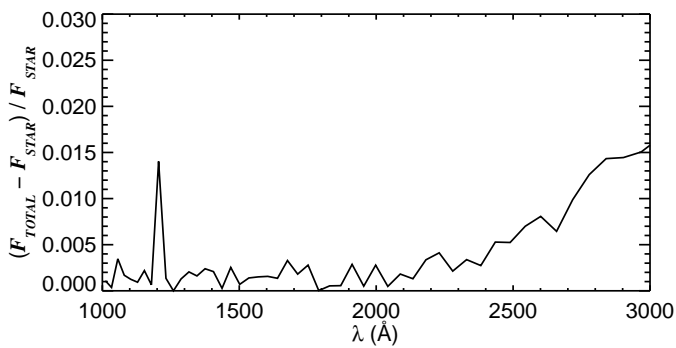}
      \caption{The model fraction of the disk flux contribution in the 1000 -- 3000~{\AA} interval computed by dividing a model containing the disk contribution by the purely photospheric model. 
              }
\label{uv_infl}
\end{figure}

\begin{table}[htb]
\caption{\label{derived}Derived model parameters}
\centering
\begin{tabular}{cc}
\hline\hline
Parameter&Value\\
\hline
$R_{\text{e}}/R_{\text{p}}$ & 1.49\\
$R_{\text{e}}$ & 4.17~$R_{\odot}$\\
$v_{\text{rot}}$ & 396~km$\cdot$s$^{-1}$\\
$v \sin(i)$ & 270~km$\cdot$s$^{-1}$\\
$\log(g)_{\text{pole}}$ & 4.09 \\
$T_{\text{pole}}/T_{\text{eq}}$ & 1.91 \\
$T_{\text{eff,pole}}$ & 13740~K \\
\hline
\end{tabular}
\end{table}

The remaining observations were used as consistency checks for the parameters determined in the way described above. The polarization level allows us to check the consistency of $i$, $\rho_0$, and $n$ in the inner parts of the disk. The equivalent widths of the different hydrogen lines provide us with another way to check the density structure in the various parts of the disk where the emission from the respective lines originates. The final consistency checks are provided by the observed interferometric visibilities, which measure the sizes of the corresponding emitting regions, along with the interferometric phase shifts, which depend on both the velocities and densities of the corresponding parts of the disk.

  \begin{figure*}[htb]
   \centering
   \includegraphics[width=16cm]{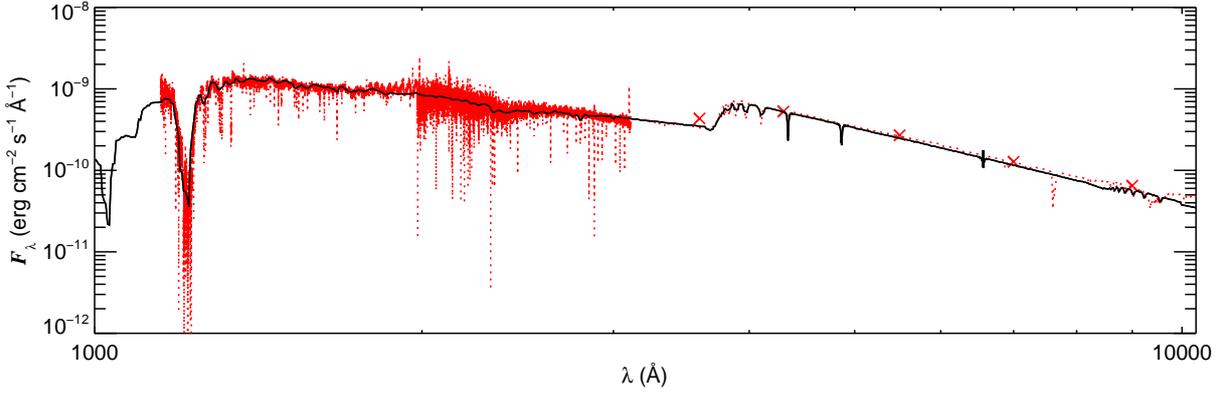}
      \caption{The parametric ($n$ = 3.5) model fit (solid black) to the average IUE and HPOL data (dotted red). The average HPOL spectrum was scaled to the $V$-band photometry. The broadband $UBVRI$ photometry is plotted with red crosses.
              }
         \label{plotsed}
   \end{figure*}

   \begin{figure*}
   \centering
         \includegraphics[width=16cm]{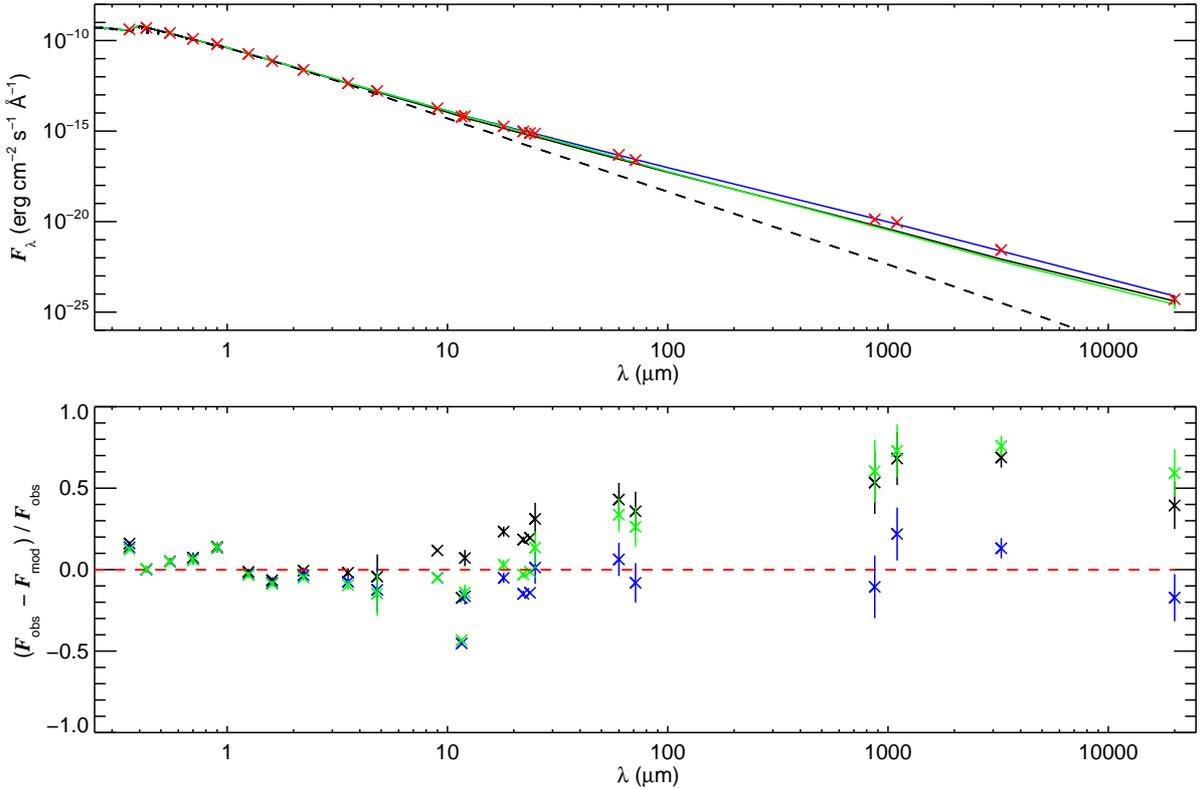}
      \caption{\textit{Upper: } parametric model with $n$ = 3.5 (black), parametric model with $n$ = 3.0 (blue), and self-consistent model (green) fit to the visual, IR, and radio SED (red plus signs). All models are plotted for $R_\text{out}$ = 35 $R_\text{p}$. The purely photospheric SED is plotted as a black dashed line. \textit{Lower: } residuals of each model fit. The error bars are plotted where available.
              }
         \label{plotir}
    \end{figure*}

Comparing the outcomes of the best-fit (\textit{\textup{parametric}} and \textit{\textup{self-consistent}}) VDD models then allows us to explore the nonisothermal effects on the observables caused by structural changes in the disk introduced with the nonisothermal density solution. Finally, we look at the evidence in our data for the disk truncation and the presence of an unseen binary companion by exploring the outer parts of the disk using the radio data and several unexpected nonhydrogen spectral features.

\subsection{SED}

\subsubsection{UV spectrum}
\label{uv}

The observed UV spectrum was found to represent  the photospheric spectrum of the central star (Fig.~\ref{uv_infl}) almost purely. Therefore the fluxes in the UV region are only influenced by the central star  and can be used to constrain its parameters. As $M$ is kept fixed and $\beta$ is a known function of $W$, the parameters that remain to be determined are $R_\text{p}$, $L,$ and $W$. While varying $L$ in the interval 160 -- 240~$L_\odot$ shifts the flux in the whole UV interval mostly in magnitude, varying $R_\text{p}$ has a much larger effect on $T_\text{pole}$ and hence on the spectral slope. With $W$ (and the corresponding $\beta$) fixed at the value determined from the analysis of visual polarimetry (Sect.~\ref{pol}), and $i$ fixed at the value determined from fitting AMBER interferometry (Sect.~\ref{amber}), $R_\text{p}$ and $L$ can be determined by fitting the UV spectral slope and the UV flux level, respectively. Out of the grid of models computed, the resulting best-fit parameters are $R_\text{p}$ = 2.8~$R_{\odot}$ and $L$ = 185 $L_{\odot}$. The model reproduction of the SED structure in the 1000 -- 10000~{\AA} interval is shown in Fig.~\ref{plotsed} (as the disk contribution is less than 3\% in these parts on average, the outcomes of the two VDD model formalisms are identical).

\subsubsection{Optical and IR SED}
As discussed in the Introduction, the size of the disk pseudophotosphere grows with wavelength. Consequently, the near-IR regions are useful to probe the inner parts of the disk and constrain the disk base density $\rho_0$, while the SED structure at longer wavelengths traces the density falloff exponent $n$. The resulting $\rho_0$ from the best-fit parametric model is 2.0 $\times$ 10$^{-12}$~g$\cdot$cm$^{-3}$; the implications for $n$ in the light of the whole SED structure are discussed below.

\begin{figure}
   \centering
         \includegraphics[width=\hsize]{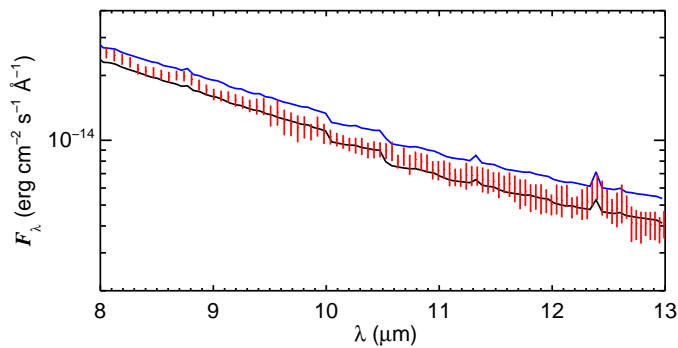}
      \caption{The parametric model with $n$=3.5 (black) and $n$=3.0 (blue) mid-IR flux predictions compared to the MIDI $N$-band photometry.
              }
         \label{midi}
    \end{figure}

The parametric model with a single value of $n$ throughout the whole disk does not seem to provide an entirely satisfactory description of the density profile. In Fig.~\ref{plotir} we see that the parametric model with $n$ = 3.5 reproduces the SED well only until $\sim$15~$\mu$m, while it underestimates the far-IR and radio fluxes by about 40--70\%. A shallower density profile with $n$ = 3.0 (and the same $\rho_0$) reproduces the SED longward of $\sim$15~$\mu$m much better at the expense of the mid-IR fluxes, which become slightly overestimated by about 20\%, as may also be seen  in Fig~\ref{midi}, which shows the comparison of the parametric model to the $N$-band MIDI photometry.

The self-consistent model SED is very similar to the parametric $n$ = 3.0 model at wavelengths up to $\sim$15~$\mu$m, showing only slight effects of nonisothermality in the inner, central parts of the disk, which are consistent with the predictions: The temperature initially decreases, which leads to a shallower radial density profile in the inner parts of the disk \citep[see Fig.~4 of][]{hdust2}. At longer wavelengths, the self-consistent model approaches the $n$ = 3.5 model, as the disk temperature rises and becomes roughly isothermal in its more extended regions. 

Overall, we find that neither the parametric model with a single $n$, nor the self-consistent model, are able to perfectly reproduce the observed SED structure, which suggests that the disk has a more complicated density profile than a simple power law, or steady decretion. We discuss the density structure further when we compare the model to the remaining observables, and we return to the issue of nonisothermality in Sect.~\ref{comp}. 

  \begin{figure}
   \centering
   \includegraphics[width=\hsize]{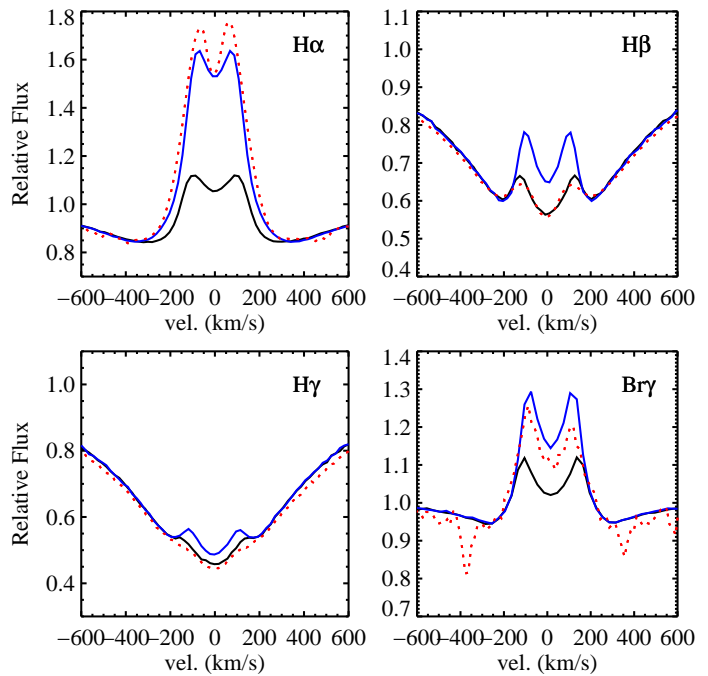}
      \caption{Hydrogen line profiles for the parametric model (solid lines) with $n$ = 3.5 (black) and $n$ = 3.0 (blue) compared to the averaged observed hydrogen line profiles (dotted red lines). The self-consistent model (not shown) is almost identical to the $n$ = 3.5 parametric model. While H$\beta$ and H$\gamma$ are reproduced well with $n$ = 3.5, H$\alpha$ and Br$\gamma$ need $n$ = 3.0 for their emission to reach the observed values.
              }
         \label{plot_line_n}
   \end{figure}

\subsection{The hydrogen emission line profiles}

  \begin{figure}
   \centering
   \includegraphics[width=\hsize]{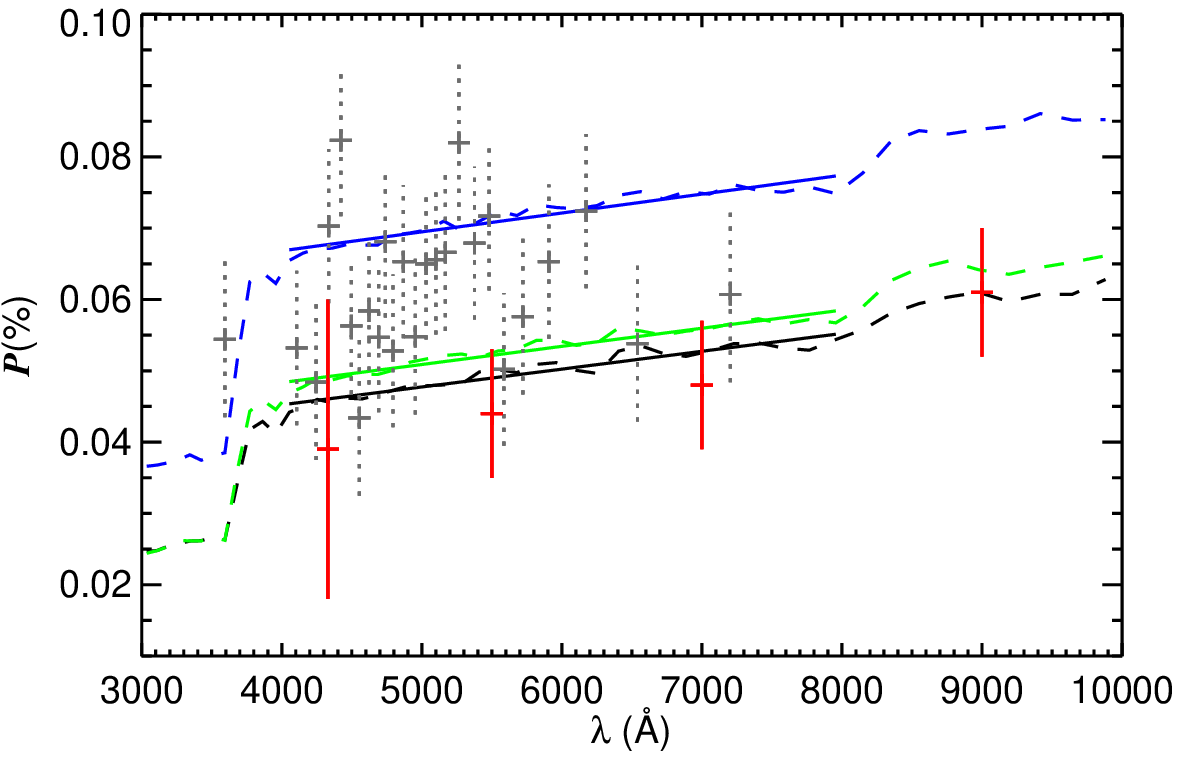}
      \caption{The spectral shapes of polarization (dashed lines) overplotted with their linear fits (solid lines) in the Paschen continuum: parametric model with $n$ = 3.5 {(black)}, parametric model  with $n$ = 3.0 {(blue),} and self-consistent model (green). The OPD measurements are plotted in red and for comparison we show the binned 1991 HPOL measurement (gray plus signs with dotted error bars).
              }
\label{plot_pol_bestfit}
\medskip
   \centering
\includegraphics[width=\hsize]{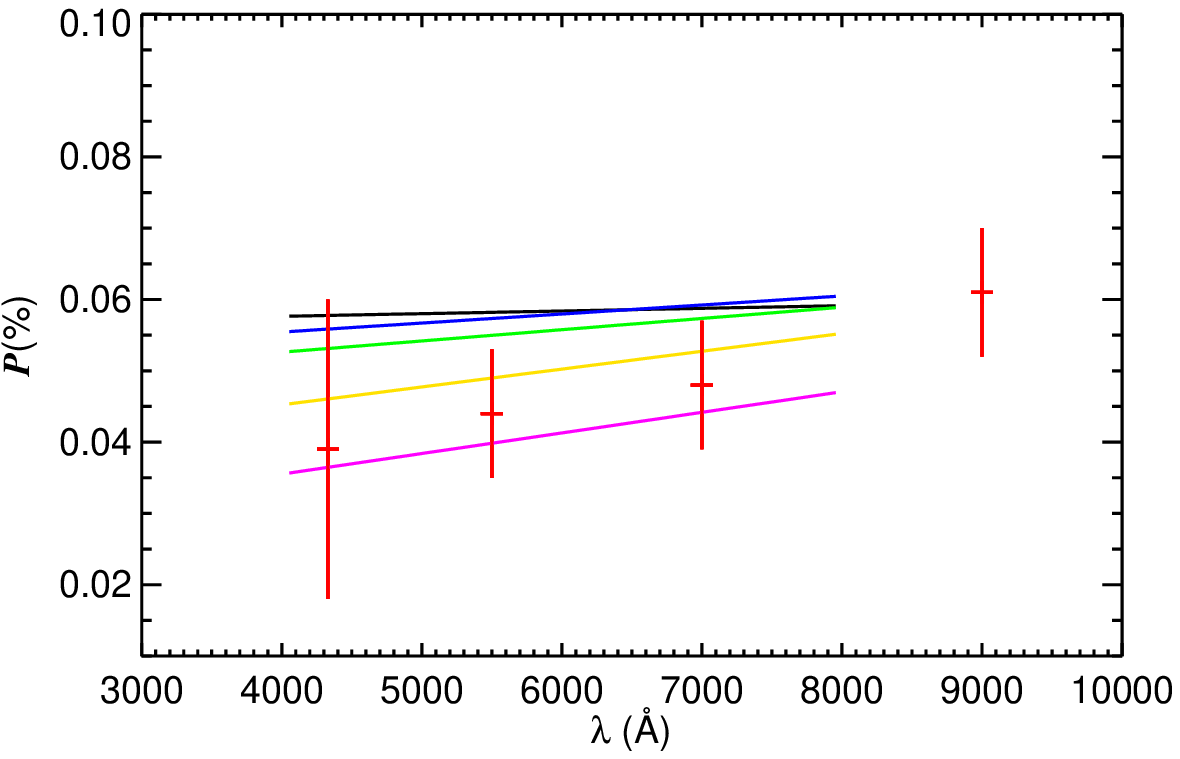}
      \caption{The influence of $W$ on the model spectral slope in the Paschen continuum for the parametric model with $n$ = 3.5. The plotted models are for $W$ = 0.80 (black), $W$ = 0.95 (blue), $W$ = 0.97 (green), $W$ = 0.99 (yellow), and $W$ = 1.0 (purple). The average of OPD measurements in the $BVRI$ filters is plotted in red.
              }
         \label{plot_pol_W}
   \end{figure}

   \begin{figure*}
   \centering
   \includegraphics[width=\hsize]{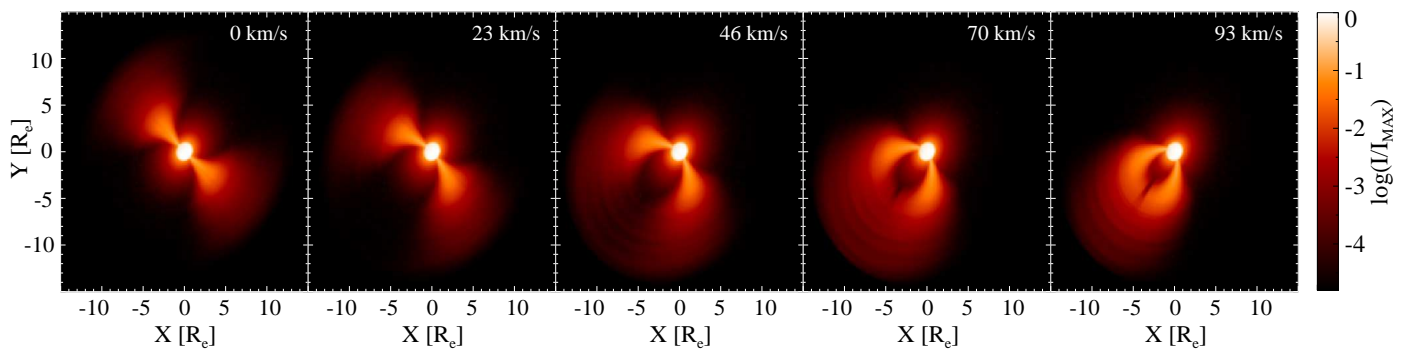}
      \caption{Synthetic images of the isovelocity regions across the Br$\gamma$ emission line. The zero velocity corresponds to the line center.
             }
         \label{betcmi_brg}
   \end{figure*}

The overall shape, equivalent width, FWHM, and the ratio of maximum to continuum intensity of the observed H$\alpha$ line profile is in  reasonable (with respect to the used methods) agreement with the historic measurements from 1950 to 1991 \citep{slettebak_reynolds, slettebak, andrillat,  hanuschik}. The hydrogen lines H$\alpha$, H$\beta$, H$\gamma,$ and Br$\gamma$ all show evidence of a double-peaked emission component. Using the best-fit parametric and self-consistent models, we calculated both line profiles and images to provide an additional consistency check of the disk density structure, as the higher Balmer lines originate progressively closer to the star. The line profile results (Fig.~\ref{plot_line_n}) clearly show that a model with a single power law describing the radial density falloff is unable to simultaneously fit all four lines. While H$\beta$ and H$\gamma$, which originate in the inner regions of the disk, are reproduced well with $n$ = 3.5, H$\alpha$ and Br$\gamma$, which come from much more extended parts of the disk, require $n$ = 3.0 to attain the observed level. 

The outcome is similar to what was suggested by the SED analysis. A steeper falloff (higher $n$) is needed in the inner parts, where H$\beta$, H$\gamma$, and near-IR continuum (but also the visual polarization, see the following section) originate, while in the more extended parts, where H$\alpha$, Br$\gamma,$ and far-IR emission come from, a shallower density profile is needed to increase the size of the optically thick emitting area and hence the line equivalent width. This again indicates that a single power law throughout the disk is not a suitable description of the density profile. The outcome of the self-consistent model for these four lines is almost identical to the parametric model with $n$ = 3.5.

For very optically thick emission lines, such as H$\alpha$, the line photons are absorbed and re-emitted many times within the Sobolev zone before they escape. This diffusion process greatly increases the probability that some fraction of the H$\alpha$ line photons are scattered by  free electrons, which have much larger thermal velocities than the hydrogen lines. To simulate this effect, we thermally broadened a fraction, $f = 0.6$, of the H$\alpha$ photons by convolving those photons with a Gaussian of width $v_\text{es}$ = 300~km$\cdot$s$^{-1}$. The remaining fraction ($1-f$) of the H$\alpha$ photons are left unbroadened. The values of $f$ and $v_\text{es}$ were chosen empirically to best reproduce the wings of the observed H$\alpha$ profile. The electron thermal velocity at $T_\text{k} = 6500~K$ is 310~km$\cdot$s$^{-1}$, which is roughly what we determined empirically for $v_\text{es}$. The remaining Balmer lines (H$\beta$, H$\gamma$ and Br$\gamma$) are much weaker, so we applied no electron scattering broadening to those lines.

\subsection{The optical polarization}
\label{pol}

The HPOL measurements show a level of variability, which is hard to quantify due to their large errors and scarce temporal coverage throughout the years 1990 -- 2005. They are consistent though with a very low level of polarization in the range of 0.01 -- 0.10\%, suggesting no large-scale mass injection or dissipation episodes have occured over the past few decades. The OPD data covering approximately the last four years show a polarization level of $\sim$0.02 -- 0.07\% in the \textit{BVRI} Johnson filters, with less scatter compared to the HPOL data and also very little temporal variation (Table~\ref{onlopd_polarimetry}, available electronically only). 

\begin{figure*}
\begin{centering}
\tikzsetnextfilename{betCMi}   
\input{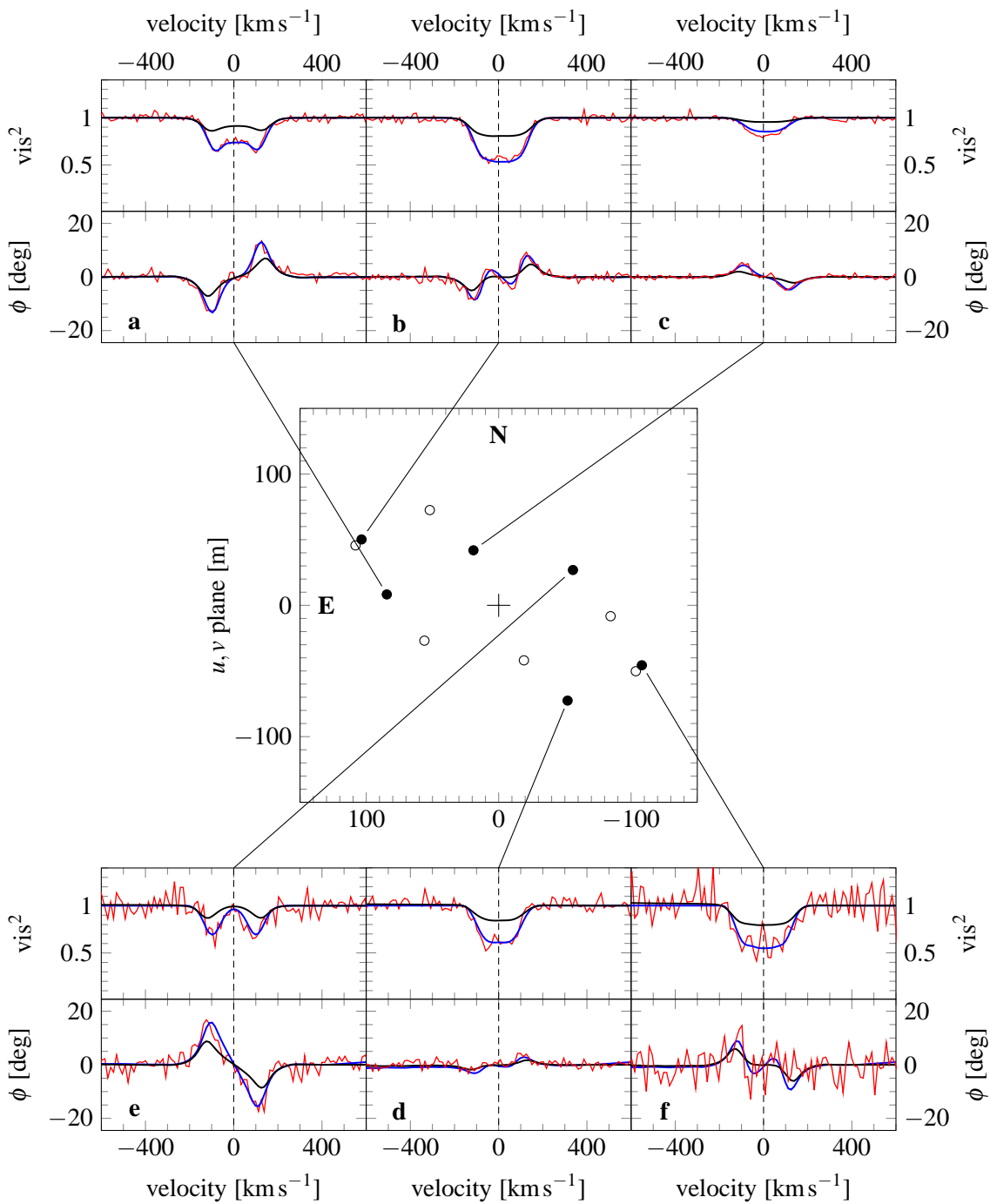}             
\caption{The parametric model with $n$ = 3.5 (black) and $n$ = 3.0 (blue) fits to the visibilities and phases measured by AMBER (red). The outcome of the self-consistent model is identical to the $n$ = 3.5 parametric model. In the middle panel, the filled circles are the $uv$-plane positions of the observations, while the open circles are their conjugates.}\label{amber_plot_n}
\end{centering}
\end{figure*}

For the modeling purposes, we used the averaged values of the OPD measurements. Despite the uncertainties being quite large, the data clearly show a positive spectral slope of the polarization level in the Paschen continuum (see Fig.~\ref{plot_pol_bestfit}). This is  an unusual feature and is not seen in any of the theoretical polarization spectra shown by \citet{haubois2}; in that work, low-density models have either flat spectra, indicative of the prevalence of Thomson scattering in the total opacity (i.e., low-density models), or negative slopes for the cases where H opacity contributes significantly to the total opacity (i.e., high-density models). In view of that, we explored which model parameter could be responsible for the positive slope. We used only the Paschen continuum polarization levels to compare the model and observed slopes to avoid the influence of Balmer and Paschen jumps. 

Increasing the rotation rate $W$ closer to the critical value was found to be the major factor causing the polarimetric slope to attain the unusual positive slope shown by the observations. The dependence of model polarization on $W$ is shown in Fig.~\ref{plot_pol_W}. Only rotation rates higher than about 0.90 produce a positive polarimetric slope, while rotation rates higher than $W = 0.98$ are required to reproduce the observations within the error bars. The dependence of the  polarimetric slope on the rotation can be probably explained by the fact that the flux constrast between the pole and the equator gets stronger at shorter wavelengths \citep[e.g., Fig. 6 of][]{review}. Since polarization is the ratio between the scattered to direct startlight, it follows that polarization should be smaller where the flux constrast is the strongest. This effect is only valid for tenuous disks whose dominant opacity (Thomson scattering) is gray. Denser disks attain a negative slope reflecting the strong contribution of H to the total opacity. The finding of  such a fast rotation is supported by the results of \citet{saio}, who found nonradial pulsation modes (prograde sectoral $g$-modes of $m=-1$) excited in $\beta$~CMi, which they only predict to be stable  if the star rotates nearly critically.

The predictions of the adopted models with $W = 0.99,$ compared to the average of OPD measurements, are shown in Fig.~\ref{plot_pol_bestfit}. The parametric model with $n$ = 3.0 overestimates the polarization level, while using $n$ = 3.5 matches the observed level well. This further corroborates what was suggested by the SED and line profile analysis, i.e., a steep density falloff in the inner parts and a shallower profile farther on.

The polarization level is strongly dependent on the inclination angle $i$ with the maximum level observed when $i \sim70-80\degree$ {\citep{halonen}}. However, the inclination was found to be well constrained by the AMBER data (see Sect. \ref{amber}), which allows us to use the polarization level to constrain the density structure of the inner disk without a degeneracy with respect to $i$.

    \begin{figure*}
   \centering
   \includegraphics[width=\hsize]{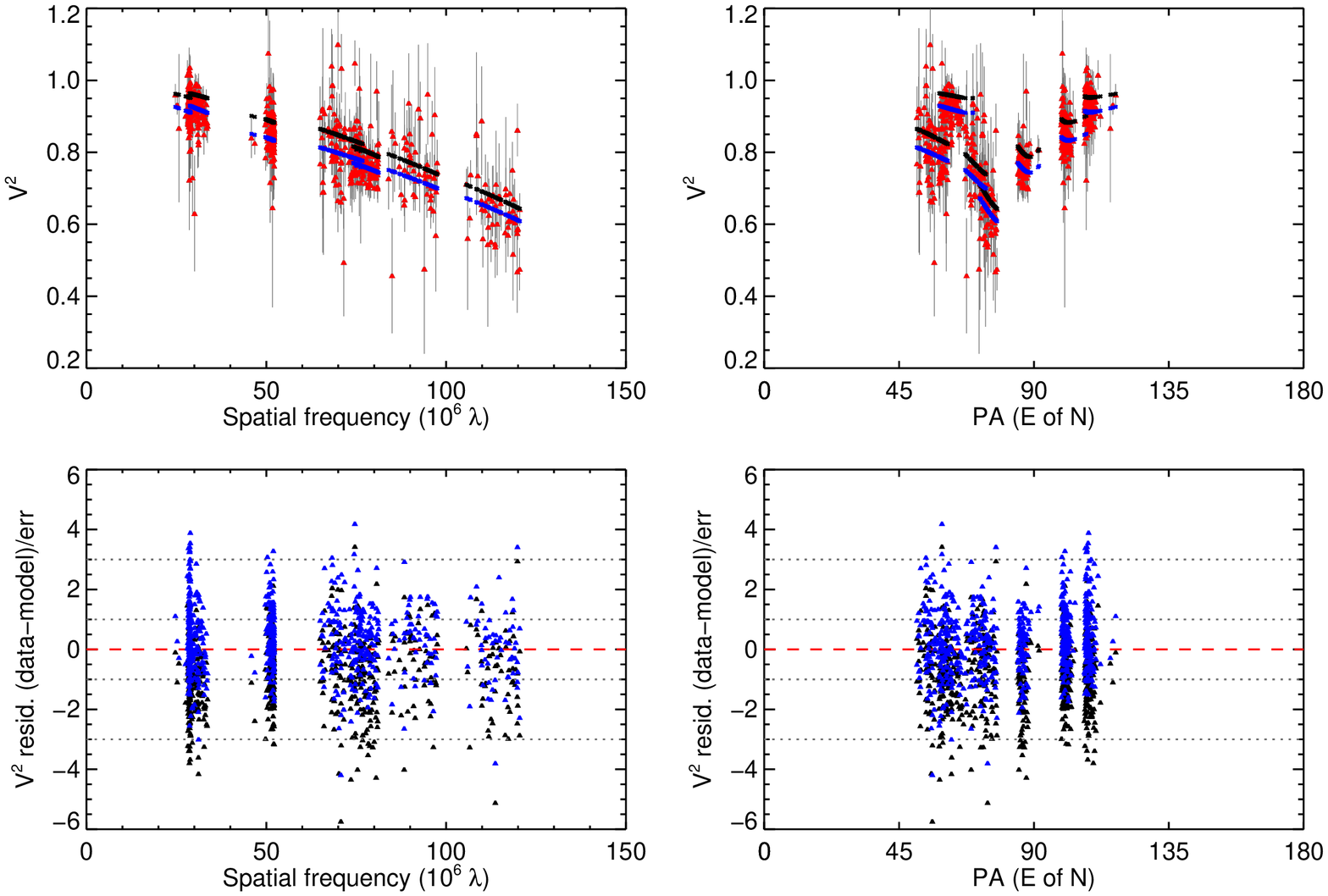}
      \caption{The NPOI H$\alpha$ channel squared visibilities (red with gray error bars) overplotted with the corresponding parametric, $n$ = 3.5, (black), and $n$ = 3.0 (blue) model predictions (compare with the upper left panel of Fig.~\ref{plot_line_n})
             }
         \label{npoi}
   \end{figure*}

\onlfig{
      \begin{figure*}
   \centering
   \includegraphics[width=\hsize]{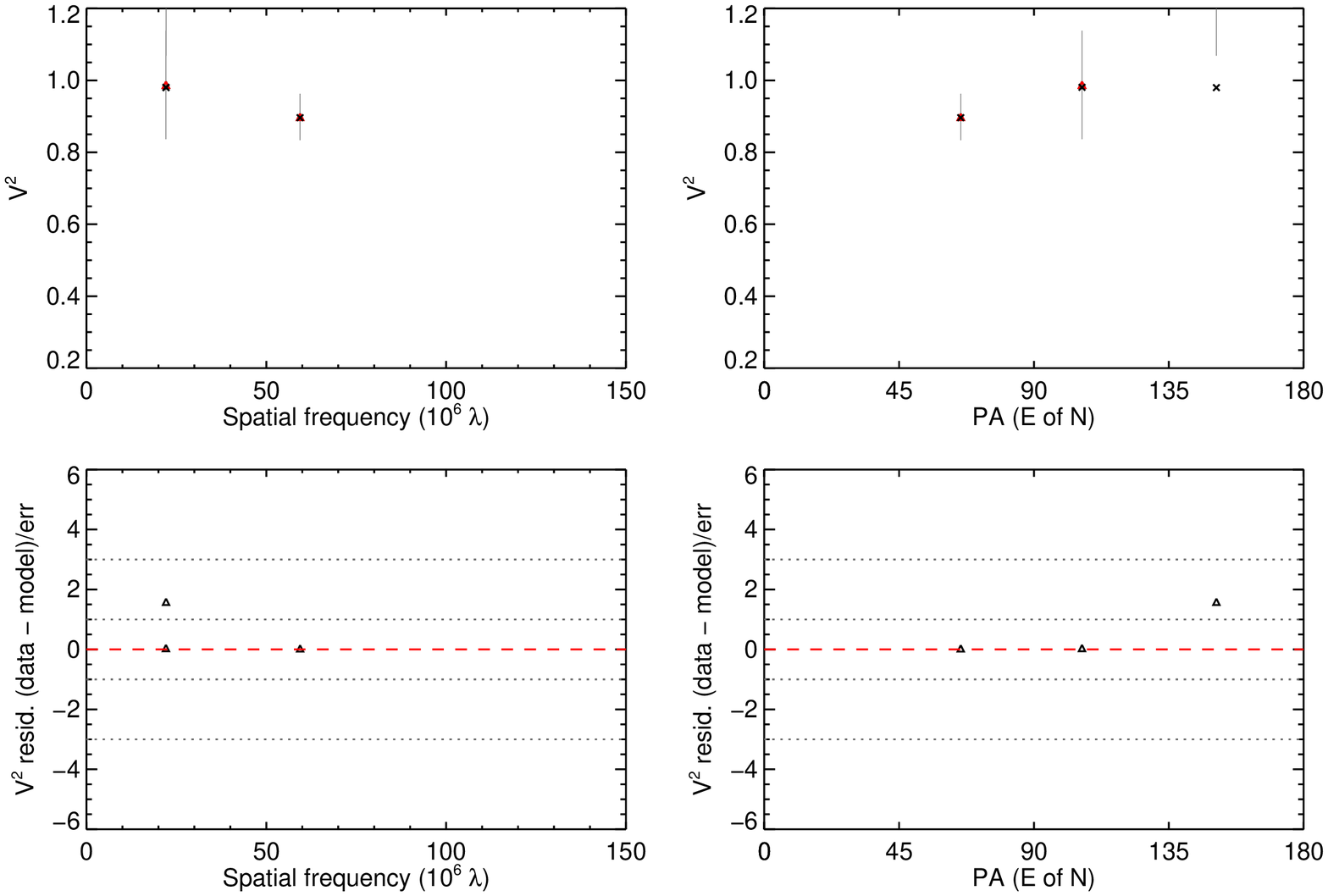}
      \caption{The FSU-A $K$-band squared visibilities (red with gray error bars) and the corresponding parametric ($n$ = 3.5) model predictions (black).
              }
         \label{prima}
   \end{figure*}
}
 
\subsection{AMBER spectrointerferometry}
\label{amber}

AMBER data allow us to study the geometry and  extension of the Br$\gamma$ emitting region. To accomplish this, we calculated the synthetic Br$\gamma$ images (several shown in Fig.~\ref{betcmi_brg}) across the emission lines, depicting the corresponding isovelocity regions. From these images we determined the predicted AMBER visibilities and phase shifts across the Br$\gamma$ line. AMBER's high spectral resolution reveals the spectrointerferometric signatures in the differential visibilities and phases that are expected for circumstellar disks. As was reported by \citet{kraus}, and discussed theoretically by \citet{faes}, phase reversals occur in the differential phase signatures of $\beta$~CMi across the Br$\gamma$ line. Using the terminology of \citet{faes}, the disk of $\beta$~CMi was observed outside the astrometric regime, which means that the ratio between the baseline length and the distance to the star exceeds $1.5 $~m$\cdot${pc}$^{-1}$. Thus the differential phases are no longer proportional to  photocenter displacement and  phase signatures are more complicated than simple S-shaped profiles usually observed for marginally resolved disks.

     \begin{figure*}
   \centering
   \includegraphics[width=\hsize]{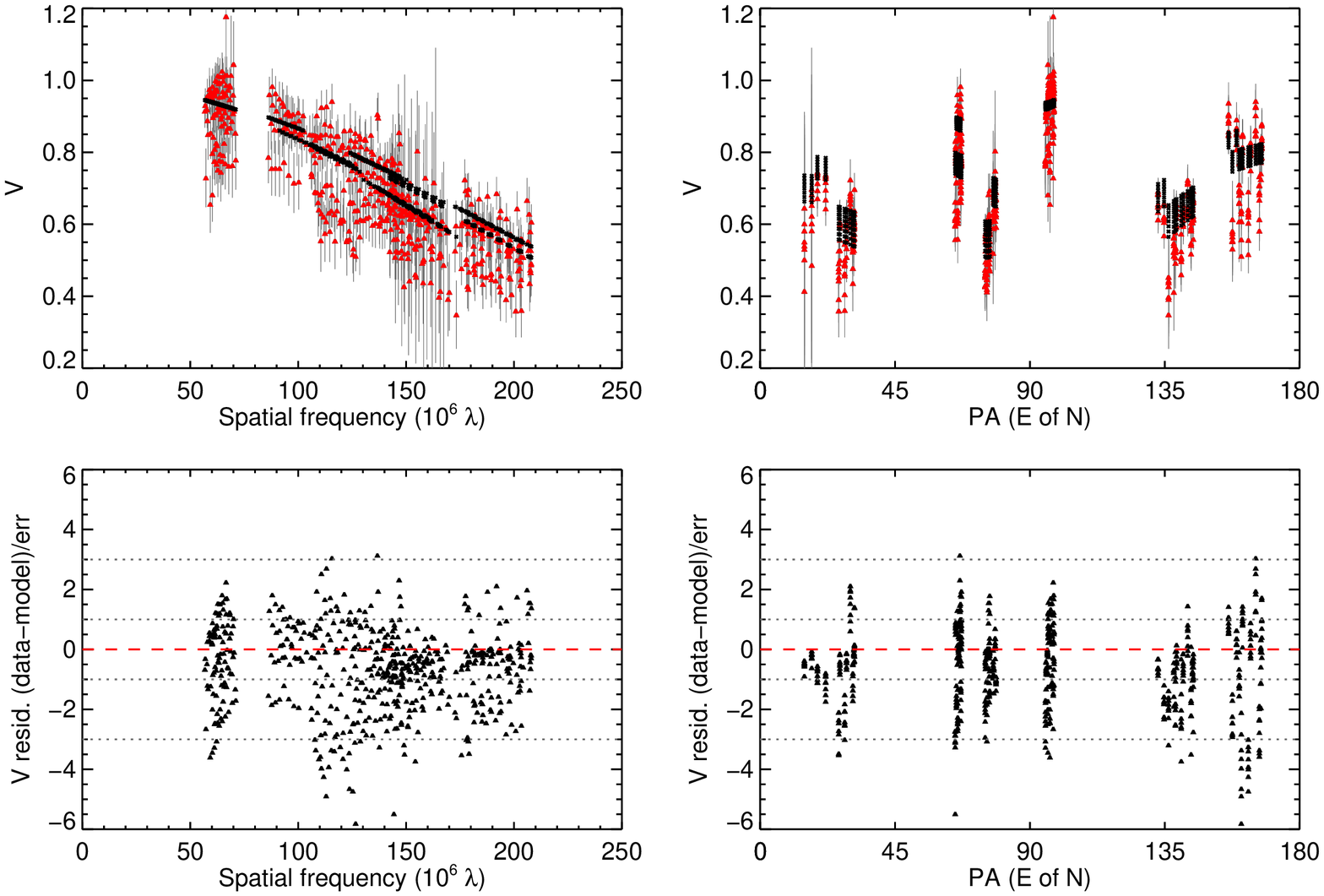}
      \caption{The MIRC $H$-band visibilities (red with gray error bars) and the corresponding parametric ($n$ = 3.5) model predictions (black).              }
         \label{mirc}
   \end{figure*}

\onlfig{
      \begin{figure*}
   \centering
   \includegraphics[width=\hsize]{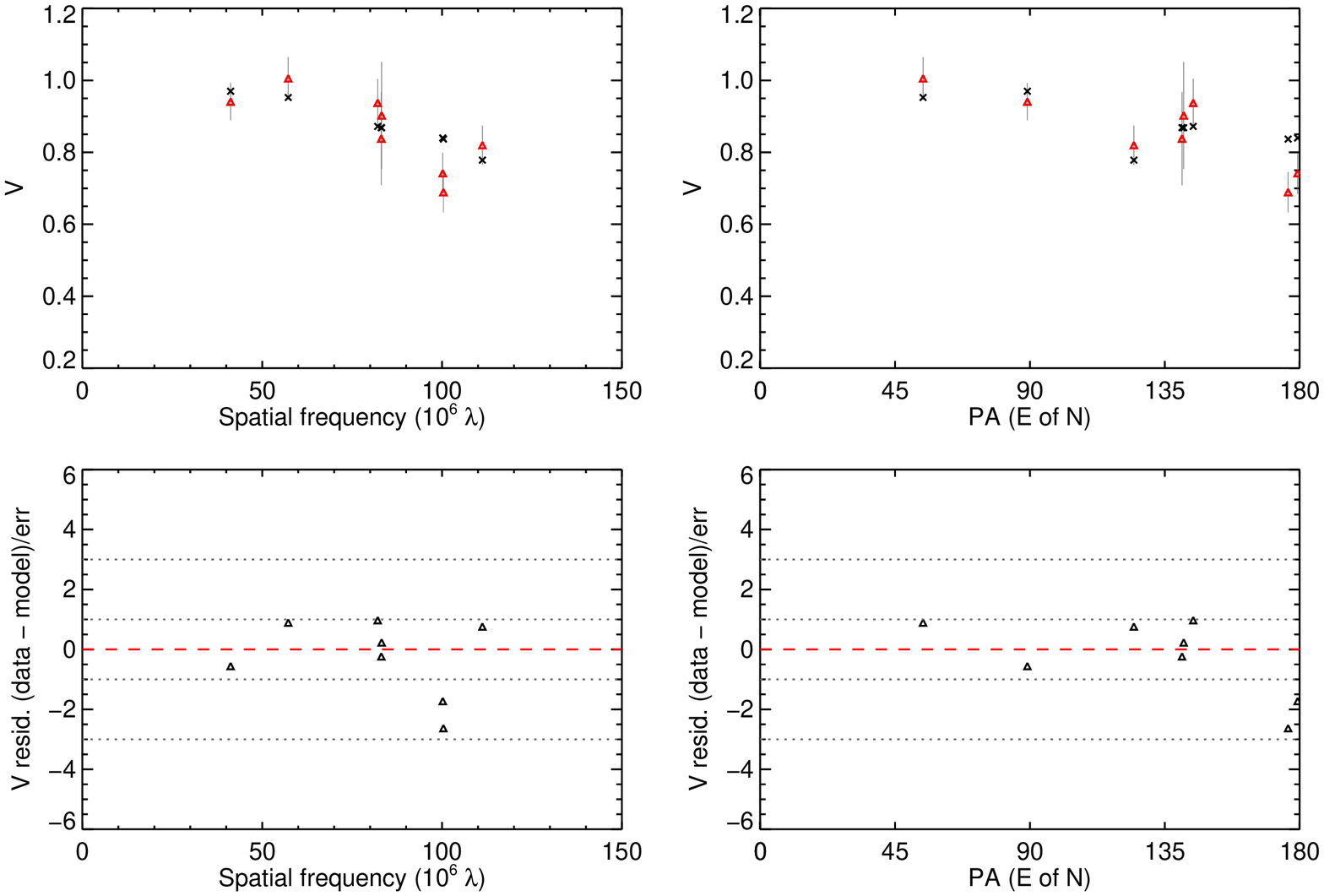}
      \caption{The CLIMB $K$'-band visibilities (red with gray error bars) and the corresponding parametric ($n$ = 3.5) model predictions (black).
              }
         \label{climb}
   \end{figure*}
}

Similar to the observables believed to originate in the more extended parts of the disk, a shallower density profile with $n$ = 3.0 is needed to reproduce the AMBER observations, while the predictions of the $n$ = 3.5 parametric model and the self-consistent model are almost identical, clearly underestimating the size of the line emitting region (Fig.~\ref{amber_plot_n}). The fit of the $n$ = 3.0 parametric model to the two sets of AMBER data is remarkable, with a reduced $\chi^2$ = 1.20. This kind of good agreement again implies that the disk velocities are Keplerian, which is strong support for the VDD model. The inclination is well constrained by the shape implied by the visibilitiy data along the various baseline orientations with the best-fit value being $i = 43^{+3\degree}_{-2\degree}$. The position angle, $PA$, of the disk major axis projected on the sky was found to be 133$^{+4{\degree}}_{-3{\degree}}$ (from north to east) in a reasonable agreement with the value of $140.0 \pm 1.7{\degree}$ found by \citet{kraus}. The average polarization position angle from the HPOL synthetic \textit{UBVRI} filter data and OPD measurements is 41.6 $\pm$ 1.8{\degree} and 50.0 $\pm$ 17.8{\degree}, respectively, in a good agreement with the interferometric $PA$, as the polarization direction is usually perpendicular to the disk major axis, but  small variations can be expected from the disk asymmetries responsible for the V/R variations. 


\subsection{NPOI interferometry}

\begin{table}
\caption{\label{interf} Model fits to the interferometric measurements.}
\centering
\begin{tabular}{cccc}
\hline\hline
Instrument&Spectral band&$\chi^2$ ($n$ = 3.5)&$\chi^2$ ($n$ = 3.0)\\
\hline
NPOI & $R$ (H$\alpha$)& 2.08 & 1.40\\
MIRC & $H$ & 2.27 & 2.08\\
FSU-A & $K$ & 0.84 & 0.91\\
AMBER & $K$ (Br$\gamma$)& 3.90 & 1.20\\     
CLIMB & $K$' & 1.58 & 1.45\\     
MIDI  & $N$ & 1.80 & 1.91\\
\hline
\end{tabular}
\tablefoot{AMBER and NPOI, for which extended line emission plays a major role, are clearly better reproduced by the $n$ = 3.0 model. The MIDI, CLIMB, and MIRC data, for which we would expect a better fit by the $n$ = 3.5 model, are also slightly better reproduced by the $n$ = 3.0 model, but not much weight should be put on these results given the overall low quality of the respective data.
}
\end{table}

 
The NPOI data, which contain the H$\alpha$ line emission in its 15~nm-wide channel, provide an important consistency check of the modeling of the H$\alpha$ line profile, which could only be reasonably reproduced  with a shallow, $n$ = 3.0, radial density falloff. Indeed, the NPOI visibilities are better reproduced with the $n$ = 3.0 model ($\chi^2$ = 1.40), revealing a larger H$\alpha$ emitting region than expected from the $n$ = 3.5 model ($\chi^2$ = 2.08). The predictions of the two models are compared with the data in Fig.~\ref{npoi}. We used an average angular radius of the central star to calibrate all the $PA$s, while the proper approach would be to use the exactly corresponding angular radii to calibrate the individual $PA$s. However, we expect that this would bring only minor improvement of the overall good fit.

\subsection{FSU-A, MIDI, PIONIER, and CHARA interferometry}

The remaining interferometric measurements are either very scarce in their $uv$-plane coverage or too noisy to provide a reliable consistency checks of the best-fit model parameters. Nevertheless, the model reproduces these data reasonably well (Table~\ref{interf}).


\subsubsection{VLTI}

The fit to the three FSU-A $K$-band visibilities is shown in Fig.~\ref{prima} (available electronically only). The MIDI data covering the $N$-band spectral region are consistent with an unresolved disk, i.e., values of visibilities $\sim$1 within 1--2$\sigma$ in a good agreement with both the $n$ = 3.5 and $n$ = 3.0 parametric models (figure not shown). The only usable outcome of the PIONIER observations was a closure phase measurement consistent with zero ($\pm$ 1.5{\degree}), which eliminates any companion up to 5\% flux contrast in the 1.5-50~mas separation range.

\subsubsection{CHARA}

Despite the MIRC data having the best $uv$-plane coverage and the longest projected baselines of all the interferometric measurements presented, significant scatter of the data, often of measurements taken during the same night for almost identical baselines prevented the data from being of much use (Fig.~\ref{mirc}). The cloud of residuals is not centered on zero, but below, i.e., the observations often show a lower fringe contrast than predicted. Diminished fringe contrast could be explained by observing issues and/or weather conditions that were not accounted for.

Finally, the model comparison to the $K$'-band CLIMB data is plotted in Fig.~\ref{climb} (available electronically only).

\subsection{Nonisothermal effects on the observables}
\label{comp}

So far, we have mostly focused on presenting the results of the parametric model as it allows us to test alternative density profiles. In this section, we compare the outcomes of the \textit{\textup{parametric}} model with $n$ = 3.5 and the \textit{\textup{self-consistent}} model and discuss the nonisothermal effects on both the disk structure and model observables that arise when the density structure is self-consistently calculated together with the temperature structure. 

\begin{figure}
   \centering
   \includegraphics[width=\hsize]{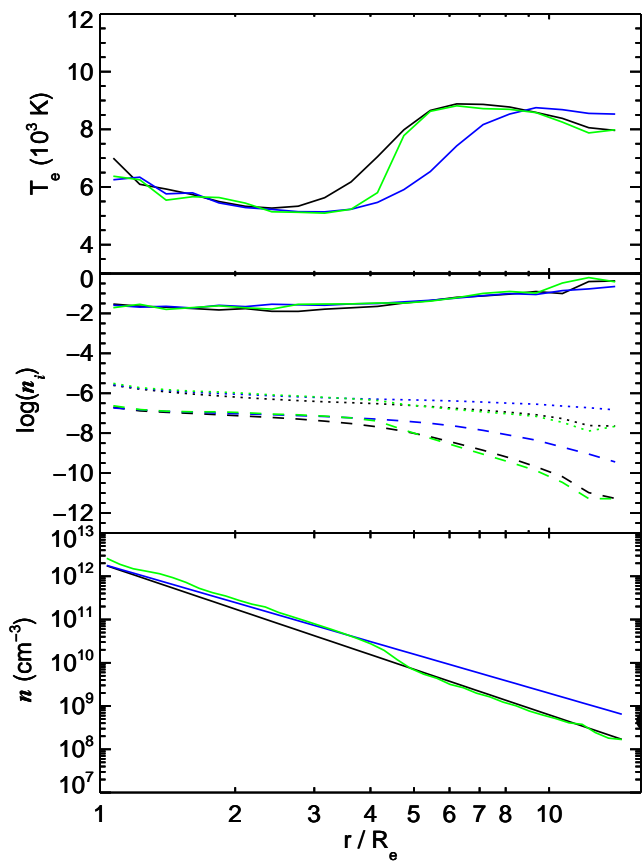}
      \caption{Electron temperature ($T_\text{e}$), hydrogen level populations (log($n_\text{i}$)), and number density ($n$) structures along the disk midplane of the parametric model with $n$ = 3.5 (black), $n$ = 3.0 (blue) and the self-consistent (green) model. The middle panel contains the $n_1$ (solid), $n_2$ (dotted), and $n_3$ (dashed) level populations.
             }
        \label{disk}
   \end{figure}

The temperature, level populations, and density structures of the two models along the disk midplane are plotted in Fig.~\ref{disk}. The temperature is found to decrease up to about 3~$R_\text{e}$, then to steeply rise, and from about 5~$R_\text{e}$ further the disk becomes isothermal and almost neutral. Introducing the self-consistent solution leads to a slight compression of the disk in the vertical direction, while the radial density structure changes as expected from the temperature structure, i.e., the falloff is shallower where the temperature gradient is negative and steeper where the temperature gradient is positive \citep{hdust2}. This has an effect on the near-IR SED structure, as  shown in Fig.~\ref{plotir}, and also on the polarization level, which slightly increases because of enhanced density in the inner parts. 

The effect of the self-consistent solution on the density structure and  SED is opposite to what was inferred from the SED and spectral line analysis, a steep density falloff followed by a shallower falloff. This appears to indicate that some mechanism, other than nonisothermality, is modifying the disk structure. Note, however, that the non-isothermal effects on the observables are small, as expected for a low-density disk around a late-type B star \citep{hdust2}. The presence of a possible unseen binary companion would most likely have stronger effects on the disk structure, and would prevent us from finding a global improvement of the model fit to the observations when making the model fully self-consistent. 


It is also interesting to note that the model Br$\gamma$ visibilities and phases are identical for the parametric model with $n$ = 3.5 and the self-consistent model, which means that while the density structure changes locally, the outcome for the whole Br$\gamma$-emitting region for these two models is the same.
We next turn to the evidence in our data for the presence of an unseen binary companion.

\subsection{Estimating the disk size from radio fluxes}
\label{s4.3.3}

  \begin{figure}
   \centering
   \includegraphics[width=\hsize]{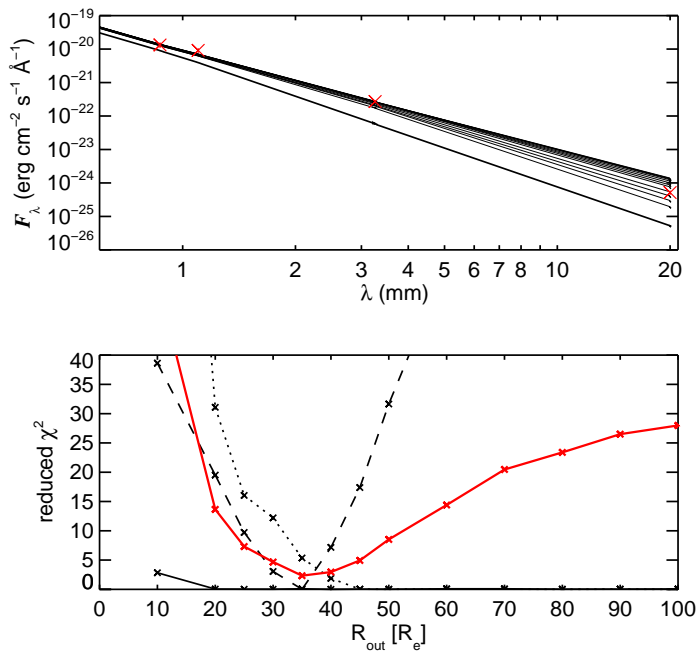}
      \caption{\textit{Upper}:  sub-mm to cm SED structure of the parametric model ($n$ = 3.0). The red crosses correspond to the observed fluxes from APEX (870~$\mu$m), JCMT (1.1~mm), CARMA (3.265~mm), and VLA (2~cm). The plotted parametric model (with $n$ = 3.0) SEDs (black solid lines) are for $R_\text{out}/R_\text{e}$ = 10; 20; 25; 30; 35; 40; 45; 50; 60; 70; 80; 90, and 100 from bottom to top. \textit{Lower}:  reduced $\chi^2$ dependence on $R_\text{out}$ for APEX (solid), CARMA (dotted) and VLA (dashed). The combined $\chi^2$ for all four observations is plotted as a red solid line.
              }
         \label{plotir_radii}
   \end{figure}

  \begin{figure*}
          \centering
  \includegraphics[width=14cm]{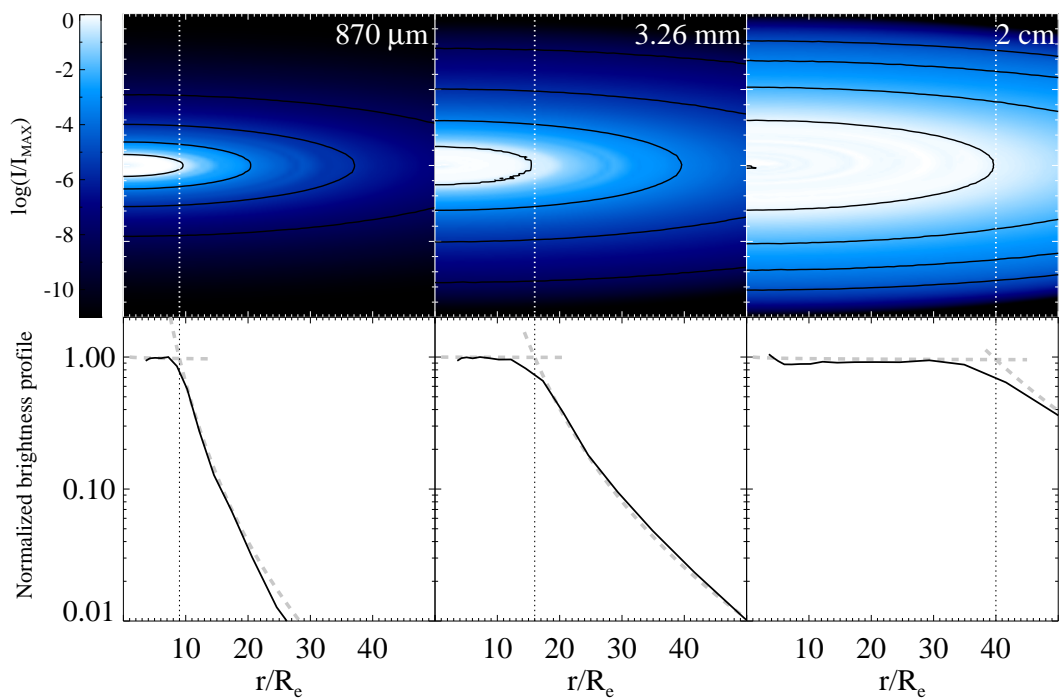}
      \caption{Synthetic images computed with {\ttfamily HDUST} (upper panels) and their respective radial brightness profile curves (lower panels), at $870\,{\rm{\mu}m}$, $3.26$~mm, and $2$~cm. In all  cases, the maximum disk emission was normalized to  unity. Overplotted to the images are the contours of same flux, at $\log(I/I_{\textrm{MAX}})=$ -0.2, -1.5, -2.5, and -3.5 (which correspond to 63\%, 3\%, 0.3\%, and 0.003\% of the disk maximum brightness, respectively). The transition between the optically thick region (pseudophotosphere) and the diffuse part of the disk is indicated by a vertical dotted line over the plots, as estimated from the extrapolation of the distinct brightness profile regions (gray dashed curves).
              }
         \label{betcmi_radio}
   \end{figure*}

Contrary to the near-IR, which probes the inner disk, the radio fluxes probe the outermost reaches of the disk. For an isolated disk, the disk can grow to the photoevaporation radius. However, if the disk is physically truncated by a binary, the pseudophotosphere effective radius eventually reaches the disk physical size and the disk flux excess no longer increases with wavelength. Consequently, the SED presents an inflection at the wavelength where $\overline{R}_\lambda\approx R_\text{out}$, and follows the photospheric spectral slope at longer wavelengths \citep[see Fig.~11 of][submitted, for an illustration]{vieira}. Therefore, the observation of this kind of transition in the SED slope can place important constraints on the disk physical size.

Under the assumption that the shallower density profile needed to reproduce the mid- to far-IR SED structure also holds  throughout the radio emitting region, we tested which disk radii reproduce best the mm and cm SED structure. In the upper panel of Fig.~\ref{plotir_radii}, we plot the sub-mm to cm SED for models with different $R_\text{out}$ and, in the lower panel, the corresponding reduced $\chi^2$. The APEX and JCMT fluxes are already reproduced  for $R_\text{out}$ = 20~$R_\text{e}$ and increasing $R_\text{out}$ beyond this value does not lead to further enhancement of the model flux. The physical explanation for this is that the size of the pseudophotosphere at these wavelengths is about 10-20~$R_\text{e}$, so increasing the disk size beyond this value does not result in an increase in the disk emission. The behavior at CARMA wavelength is similar, but $R_\text{out}$ of at least 40~$R_\text{e}$ is needed. At 2~cm, however, the situation is different. While the lowest residual between the model and observed flux is seen for $R_\text{out}$ = 35~$R_\text{e}$, a higher value of $R_\text{out}$ overestimates the observed flux. The best-fit value of $R_\text{out}$ is found to be 35$^{+10}_{-5}$~$R_\text{e}$. 

    \begin{figure*}
   \centering
   \includegraphics[width=\hsize]{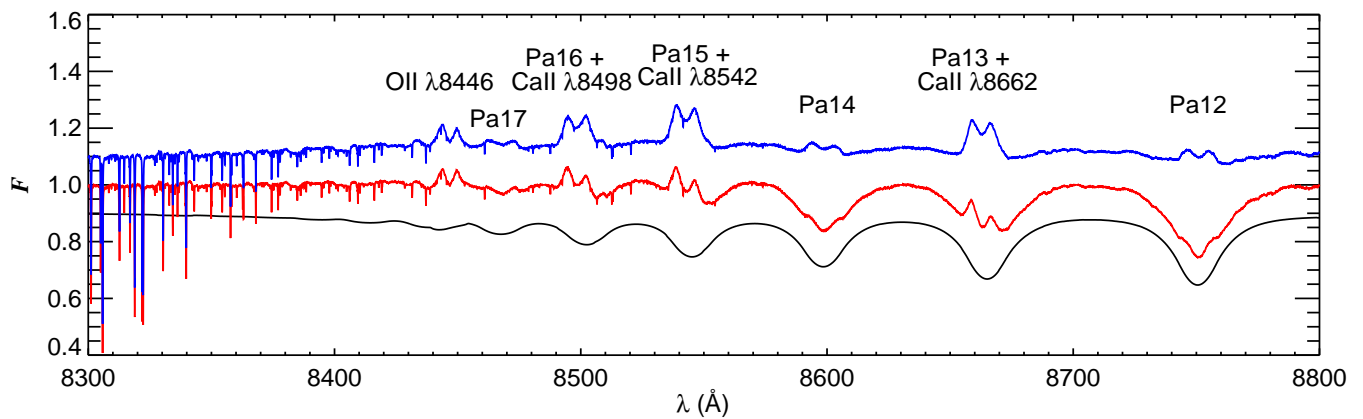}
      \caption{ Ca triplet spectral feature. The BRUCE model photospheric spectrum (black) is plotted alongside the averaged normalized spectrum observed by ESPaDOnS (red) and the difference between the two, which represents the pure emission component (blue). Both the observed and  pure emission spectrum show that the Paschen lines Pa16, Pa15, and Pa13 blended with calcium emission CaII~$\lambda$8498, 8542, and 8662, respectively, are much stronger than the emission from the unblended Paschen lines.
             }
         \label{Ca_triplet}
   \end{figure*}

    \begin{figure*}
  \centering
   \includegraphics[width=\hsize]{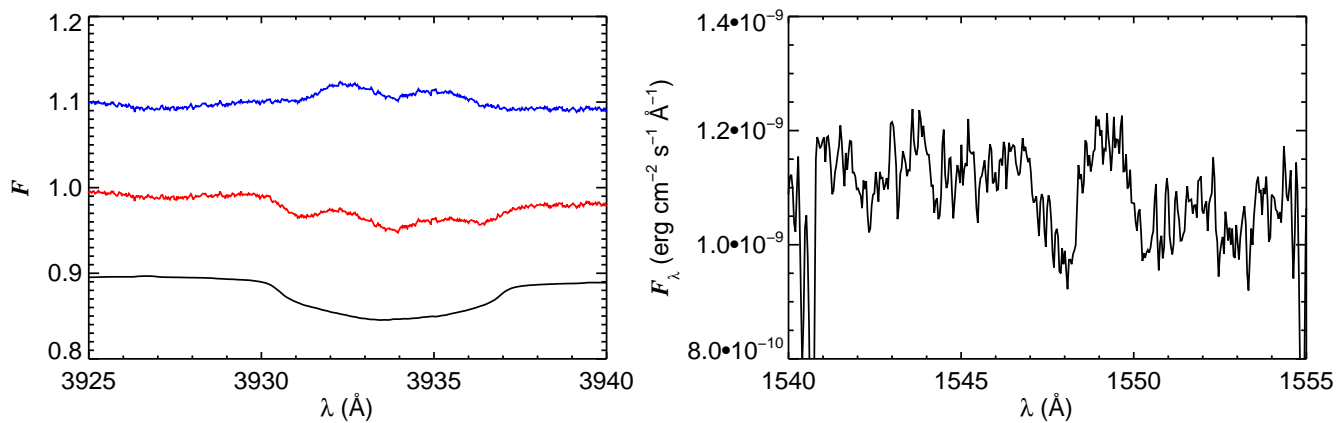}
      \caption{\textit{Left: }  CaII~$\lambda$3934 line. The purely photospheric model absorption (black) is plotted alongside the normalized ESPaDOnS spectrum (red) and their difference is plotted in blue. The emission component is clearly narrower than the absorption. \textit{Right: }  CIV~$\lambda$1548 line as observed by IUE showing a clear P~Cygni profile.
                     }
         \label{Ca+cIV}
  \end{figure*}

In Fig.~\ref{betcmi_radio} we show the synthetic images and graphical representation of the brightness profiles at wavelengths corresponding to the APEX, CARMA, and VLA observations with the estimates of the corresponding pseudophotosphere sizes. Indeed, as expected from the model behavior, at the shorter wavelengths $\overline{R}_\lambda$ is between 10--20~$R_\text{e}$. However, $\overline{R}_\lambda$ at 2~cm is about 40~$R_\text{e}$, thus larger than the estimated disk size. The 2~cm results can therefore be interpreted as a clear sign of truncation by an unseen binary companion, however,  this conclusion rests on the reliability of a single non-contemporaneous data point.

\subsection{Spectral features of other elements}

To explore the additional spectral features in the interval covered by the FEROS and ESPaDOnS spectra, we used an updated version of the BRUCE3 code \citep[see][and the references therein]{bruce} to compute a purely photospheric, high-resolution spectrum for the stellar parameters listed in Table~\ref{t5}.


An unusual feature is present in both the FEROS and ESPaDOnS spectra. The calcium triplet (CaII~$\lambda$8498, 8542, and 8662~\AA) is in emission (blended with the Paschen lines Pa16, Pa15, and Pa13), but is noticeably stronger than the emission from the unblended Paschen line Pa14 (Fig.~\ref{Ca_triplet}). This effect was linked to possible binarity of Be stars by \citet{polidan}, while a correlation of the presence of this feature in a number of B subdwarf (sdB) stars with a cool companion was reported by \citet{jeffery}. The emission in this case probably originates in the very outer disk, which is possibly partially accreting onto the secondary component. 

Moreover, in the CaII~$\lambda$3934 line, the emission component is noticeably narrower than the absorption component (left panel of Fig.~\ref{Ca+cIV}), while the opposite is expected for a line originating in the fast-rotating inner disk. Therefore, similarly to the calcium triplet mentioned above, the emission seems to actually originate in the outer disk, close to a possible secondary companion.

Finally, the ultraviolet CIV~$\lambda$1548 line exhibits a P~Cygni profile with a blue edge velocity of $\sim$230~km$\cdot$s$^{-1}$ (right panel of Fig.~\ref{Ca+cIV}), which is a clear signature of a fast stellar wind. The presence of this line is not expected in this late-type B star \citep[see Fig.~11 of][]{review}. Additionally, the observed velocity is lower than is typical for late B stars. This suggests that this kind of wind might be excited by the radiation of a companion, which interacts with the outer disk of the primary Be star.


\section{Discussion on the possible binarity}
\label{binarity}

Using the radio flux measurement at the wavelength of 2~cm in combination with a well-sampled SED at shorter wavelengths, we found observational evidence suggesting truncation of the disk of $\beta$~CMi at about 35~$R_\text{e}$. The most probable mechanism that can truncate the disk at this radius is the tidal influence of an orbiting binary companion. The presence of an unseen binary companion is also suggested  by weak $V/R$ oscillations of the H$\alpha$ line and the appearance of the CaII~$\lambda$3934 emission line and the triplet CaII~$\lambda$8498, 8542, 8662. Additionally, the CIV~$\lambda$1548 line sugggests a hot, possibly subdwarf (sdB or sdO) companion. The proposed companion has to be faint, as its signatures are seen neither in the spectra nor in the SED. The detection limits from interferometric measurements set the spectral type of a main-sequence companion to be F5 or later, thus putting an upper limit of $\sim$0.5 on the mass ratio $q$.

\begin{figure*}
   \centering
   \includegraphics[width=17cm]{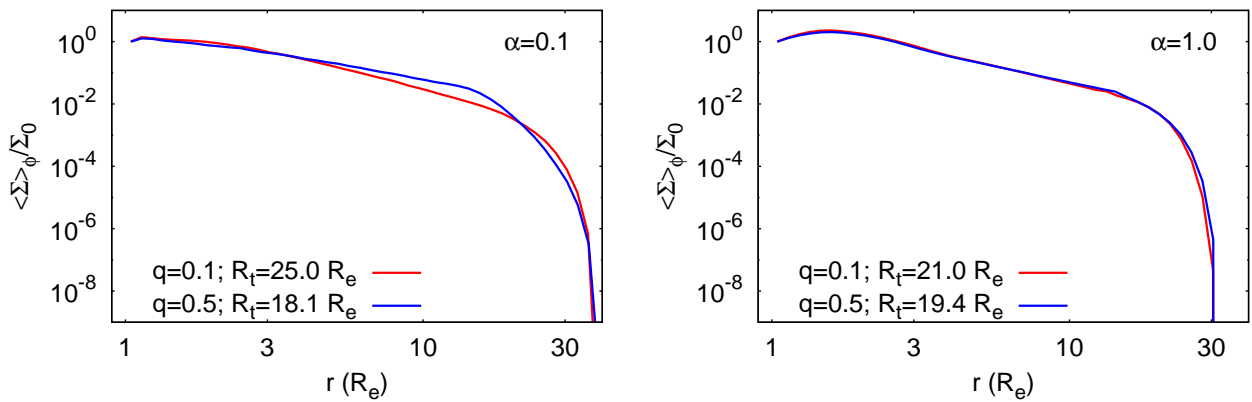}
      \caption{Results of the SPH simulations showing density profiles of isothermal disks influenced by the assumed binary companion in a circular orbit with $P=182.83\,$days, and different mass ratios, as indicated. {\it Left}: simulations for viscosity parameter $\alpha=0.1$. {\it Right}: simulations for $\alpha=1$. What is plotted is the azimuthally averaged surface density after 30 orbital periods. The truncation occurs where the density slope changes to a much steeper slope. 
             }
        \label{sph}
   \end{figure*}

To test whether a binary companion could truncate the disk at the radius estimated from the modeling, we made the assumption that the $V/R$ variations in H$\alpha$ are indeed caused by a companion, in which case the period of these oscillations \citep[$\sim$183 days,][]{folsom} can be used as an estimate of the orbital period. We then performed simulations of the binary system in a circular orbit with a smoothed particles hydrodynamics (SPH) code \citep{okazaki2002}, which assumes an isothermal disk, to determine the truncation radius, $R_\text{t}$, resulting from the presence of the possible companion. In the SPH simulation, the Be star disk is initially nonexistent. We then assume a constant mass injection rate from the equator of the star onto the equatorial plane, so that after letting the system evolve, we achieve a base density $\rho_0$ equal to its constrained value (Table~\ref{t5}) . We explored the outcome of those simulations (shown in Fig.~\ref{sph}) for different viscosity parameters, $\alpha\in\{0.1,1.0\}$, and mass ratios, $q\in\{0.1,0.5\}$. Figure~\ref{sph} shows the surface density structure of representative models after a long period of disk evolution (30 orbital periods) to ensure that the disk has reached a quasi-steady-state configuration \citep{panoglou, haubois1}. Following \citet{okazaki2002}, the truncation radius is estimated by fitting the expression
\begin{equation}
\Sigma = A\frac{(r/R_\text{t})^{-k}}{1+(r/R_\text{t})^l}
\end{equation}
to the results of the SPH simulations, where $A$, $R_\text{t}$, $k$, $l$ are constants to be fit. The resulting truncation radii range from $\sim$18~$R_\text{e}$ ($\alpha=0.1$; $q=0.5$) to $\sim$25~$R_\text{e}$ ($\alpha=0.1$; $q=0.1$), as compared to an $R_\text{out}$ of 35$^{+10}_{-5}$~$R_\text{e}$ determined from the SED fitting. A certain discrepancy is to be expected because in the {\ttfamily HDUST} model, the disk is entirely cut off at $\sim$35~$R_\text{e}$. In contrast, the truncation radius in the SPH simulations represents a point were the density falloff becomes steeper (Fig.~\ref{sph}), but the disk still exists beyond this point. This means that the size of the pseudophotosphere at 2~cm should in principle be somewhat larger than $R_\text{t}$, which would reconcile the two results. 

Out of the explored mass ratios and viscosity parameters, the $q = 0.5$ and $\alpha = 0.1$ simulation is the only one that clearly causes the density structure of the disk inward of $R_\text{t}$ to attain a shallower profile of $n = 3.0$, which is the value that was determined for intermediate part of the disk (inward of $R_\text{out}$) from the {\ttfamily HDUST} modeling. However, the steeper density profile in the very inner parts of the disk is not reproduced by the simulation, although there is a possibility that the introduction of nonisothermality into the SPH simulation might show the desired effect.

In any case, although the proposed binarity is  quite probable, this must be viewed with caution. Firstly, the evidence for the disk truncation is based only on one single measurement from more than 30 years ago. Despite the apparent stability of the disk over the last $\sim$15 years, as shown by our spectroscopic data set, and an overall stability over the last $\sim$65 years according to the literature, we cannot exclude the possibility that the mass injection rate into the disk has changed over the last decades to a degree that would destroy evidence of truncation. Secondly, it is unclear whether the complicated density structure including the inner parts is compatible with the presence of the binary companion, although the tidal interaction seems to offer the most straightforward explanation. A careful analysis of nonisothermal SPH simulations is necessary to explore if the presence of the companion offers the explanation, or if other causes are needed. Since this is beyond the scope of the present analysis, we plan to return to this issue in a follow-up paper. In that paper, we intend to study the density profile in more detail as well as execute our own analysis of the $V/R$ oscillations using all available measurements of H$\alpha$ and possibly other emission lines.

In addition to binarity, possible mechanisms for truncating the disk are photoevaporation by gas pressure and line driven radiative ablation \citep{wind1, wind2}. The former was already examined in Sect.~\ref{s3.2} and affects the disk structure only at very large radii. The latter, however, is a possibility to consider. Using the stellar wind mass-loss rate estimate of \citet{castor}, \citet{krticka} have determined that the disk wind mass-loss rate scales with the stellar wind mass-loss rate, but with values lower by more than an order of magnitude. This implies that the disk ablation produces a mass loss much lower than that of a main-sequence star. As the winds of late-type main-sequence B stars are typically weak, we conclude that for stars, such as $\beta$ CMi, wind ablation should not play a role in the truncation of the disk.



\section{Conclusions}

We tested the ability of the VDD model to reproduce arguably the largest data set of multiwavelength and multitechnique observations of a Be star ($\beta$~CMi), including the SED covering five orders of magnitude in wavelength, emission line profiles of four hydrogen lines, visual polarimetry, high-resolution, near-IR spectrointerferometry and visual, near-IR, and mid-IR broadband interferometry. Neither a single power-law structure nor steady viscous decretion appears to be an entirely adequate description of the density profile. Observables originating in the inner disk (polarization, higher Balmer lines, near-IR photometry) require the steeper radial power-law exponent $n$ = 3.5, while observables originating from more extended regions (H$\alpha$, Br$\gamma$, mid- to far- IR photometry, sub-mm/mm photometry) require the shallower $n$ = 3.0. It is shown that nonisothermal effects on the disk structure cannot account for this more complicated density behavior. Instead, an outer tidal influence most probably exerted by an unseen binary companion is needed, which remains a possibility to be investigated in detail in the future.

Analysis of  radio observations suggests truncation of the disk at the distance of 35$^{+10}_{-5}$~$R_\text{e}$ from the star. This value was compared with the truncation radius exerted by a possible binary companion with the orbital period corresponding to the period of weak $V/R$ oscillations of the H$\alpha$ line with an overall good agreement. Together with spectroscopic evidence and a nondetection by any of the other observations, either a late (F5 or later) main-sequence companion or a subdwarf (sdB or sdO) companion is favored.

We report on the diagnostic potential of the polarimetric slope in the Paschen continuum for constraining the rotation rate $W$. The observed positive slope could only be reproduced with rotation rates very close to critical, with a favored value of $W \gtrsim 0.98$. The almost purely photospheric UV spectrum allowed us to constrain the polar radius $R_\text{p} = 2.8~R_{\odot}$ and luminosity $L = 185~L_{\odot}$ of the central star. Finally, AMBER spectrointerferometry facilitated a precise estimation of the inclination angle ($i = 43^{+3\degree}_{-2\degree}$). Therefore, to conclude, these results underscore the utility of multitechnique and multiwavelength observations to  constrain the disk structure unambiguously, as well as  the use of radio observations  to study the properties of the outer parts of Be disks.


\begin{acknowledgements}

R.~K. would like to acknowledge the kind help of Dietrich Baade in the final stages of preparation of the paper. R.~K. thanks Petr Harmanec for being the observer of several spectra used in this work.

The research of R.~K. was supported by grant project number 1808214 of the Charles University Grant Agency (GA UK) and by the research program MSM0021620860 (M\v{S}MT \v{C}R). 

A.~C.~C. acknowledges support from CNPq (grant 308985/2009-5).

D.~M.~F. acknowledges support from FAPESP (grant 2010/19060-5).

R.~G.~V. acknowledges the support from FAPESP (grant 2012/20364-4).

D.~P. acknowledges support from FAPESP grant No 2013/16801-2.

M.~C. acknowledges, with thanks, the support of FONDECYT project 1130173 and Centro de Astrof\' isica de Valpara\' iso.

This work has made use of the computing facilities of the Laboratory of Astroinformatics (IAG/USP, NAT/Unicsul), whose purchase was made possible by the Brazilian agency FAPESP (grant 2009/54006-4) and the INCT-A.

Support for CARMA construction was derived from the states of California, Illinois, and Maryland, the James S. McDonnell Foundation, the Gordon and Betty Moore Foundation, the Kenneth T. and Eileen L. Norris Foundation, the University of Chicago, the Associates of the California Institute of Technology, and the National Science Foundation. Ongoing CARMA development and operations are supported by the National Science Foundation under a cooperative agreement, and by the CARMA partner universities. 

The Navy Precision Optical Interferometer is a joint project of the Naval Research Laboratory and the US Naval Observatory, in cooperation with Lowell Observatory and is funded by the Office of Naval Research and the Oceanographer of the Navy. C.~T. would also like to thank student assistants, Bryan Demapan, Brennan Kerkstra, and Sandeep Chiluka for assistance with NPOI data reductions.

Some of the data presented in this paper were obtained from the Mikulski Archive for Space Telescopes (MAST). STScI is operated by the Association of Universities for Research in Astronomy, Inc., under NASA contract NAS5-26555. Support for MAST for non-HST data is provided by the NASA Office of Space Science via grant NNX13AC07G and by other grants and contracts.      
\end{acknowledgements}


\bibliographystyle{aa} 
\bibliography{biblio.bib} 


\end{document}